\documentclass[journal]{IEEEtran}

\usepackage{graphicx}
\usepackage{enumerate}
\usepackage{cite}
\usepackage{bm}
\usepackage{color}
\usepackage{xcolor}
\usepackage{amsmath,amssymb,amsthm}
\usepackage[linesnumbered,boxruled,vlined]{algorithm2e}
\usepackage{hyperref}
\usepackage{booktabs}  
\usepackage{config}
\usepackage{flushend}
\usepackage[font=footnotesize]{caption}

\newcounter{propcounter}
\renewcommand{\thepropcounter}{\Roman{propcounter}}

\newcommand{\property}{%
  \refstepcounter{propcounter}%
  \paragraph*{Property \thepropcounter}%
}

\newcounter{lemcounter}
\renewcommand{\thelemcounter}{\Roman{lemcounter}}

\newcommand{\lemma}{%
  \refstepcounter{lemcounter}%
  \paragraph*{Lemma \thelemcounter}%
}

\makeatletter
\renewcommand{\Indentp}[1]{%
  \advance\leftskip by #1
  \advance\skiptext by -#1
  \advance\skiprule by #1}%
\renewcommand{\Indp}{\algocf@adjustskipindent\Indentp{\algoskipindent}}
\renewcommand{\Indm}{\algocf@adjustskipindent\Indentp{-\algoskipindent}}
\let\oldnl\nl
\newcommand{\nonl}{\renewcommand{\nl}{\let\nl\oldnl}}
\makeatother

\begin{document}

\title{On the Robustness of RSMA to Adversarial BD-RIS-Induced Interference}
\author{Arthur S. de Sena, \textit{Member}, \textit{IEEE},
Jacek Kibi\l{}da, \textit{Senior Member}, \textit{IEEE},
Nurul H. Mahmood, \textit{Member}, \textit{IEEE},
\\André Gomes, \textit{Member}, \textit{IEEE}, Luiz A. DaSilva, \textit{Fellow}, \textit{IEEE}, Matti Latva-aho, \textit{Fellow}, \textit{IEEE}

\thanks{Copyright (c) 2026 IEEE. Personal use of this material is permitted. However, permission to use this material for any other purposes must be obtained from the IEEE by sending a request to pubs-permissions@ieee.org.}
\thanks{Arthur S. de Sena, Nurul H. Mahmood, and Matti Latva-aho are with the University of Oulu, Oulu, Finland (email: arthur.sena@oulu.fi, nurulhuda.mahmood@oulu.fi, matti.latva-aho@oulu.fi).}
\thanks{Jacek Kibi\l{}da and Luiz DaSilva are with the Commonwealth Cyber Initiative, Virginia Tech, USA (email: jkibilda@vt.edu, ldasilva@vt.edu).}
\thanks{André Gomes is with Rowan University, USA (email: gomesa@rowan.edu).}}

\maketitle
\begin{abstract}
    This article investigates the robustness of \ac{RSMA} in multi-user \ac{MISO} systems to interference attacks against channel acquisition induced by \acp{BD-RIS}. Two primary attack strategies, random and aligned interference, are proposed for fully connected and group-connected \ac{RIS} architectures. Valid random reflection coefficients are generated exploiting the Takagi factorization, while potent aligned interference attacks are achieved through optimization strategies based on a \ac{QCQP} reformulation followed by projections onto the unitary manifold. Our numerical findings reveal that, when perfect \ac{CSI} is available, \ac{RSMA} behaves similarly to \ac{SDMA} and thus is highly susceptible to the attack, with \ac{BD-RIS} inducing severe performance loss and significantly outperforming diagonal RIS. However, under imperfect \ac{CSI}, \ac{RSMA} consistently demonstrates significantly greater robustness than \ac{SDMA}, particularly as the system's transmit power increases.
\end{abstract}

\begin{IEEEkeywords}
	Reconfigurable intelligent surface, rate-splitting multiple access, physical-layer security, multi-user \ac{MISO}.
\end{IEEEkeywords}

\acresetall

\section{Introduction}

6G introduces a range of new technologies that will positively contribute to improved service and feature offerings at the expense of greatly expanding the threat surface. \Ac{RIS} is one example of a benign technology that is anticipated to contribute to improved coverage, spectral efficiency, and sensing services, but one that has also attracted significant interest in the adversarial context~\cite{lyu2020irs,wang2022wireless,staat2022mirror,alakoca2023metasurface,huang2023disco, Sena24, Wang24,rivetti2024malicious}. Being a passive relay, \ac{RIS} can be used as a passive jammer that can adjust its reflective properties to execute an adversarial action, e.g., canceling the legitimate signal~\cite{lyu2020irs}, aiding the active jammer~\cite{wang2022wireless}, or poisoning the channel estimation process~\cite{staat2022mirror}. The latter attack is of particular severity, since channel acquisition is a critical component of advanced wireless communication systems that rely on multi-antenna technology. Here, \ac{SDMA} has been shown to be susceptible to a \ac{RIS}-induced attack that requires no or limited \ac{CSI}~\cite{huang2023disco, Sena24, Wang24,rivetti2024malicious}. However, not all multiple access protocols are equally vulnerable. Specifically, in our preliminary work on \ac{RSMA}~\cite{Sena24b}, we observed a surprising robustness of \ac{RSMA} to \ac{RIS}-induced attacks.

\begin{figure}[t]
	\centering
	\includegraphics[width=1\linewidth]{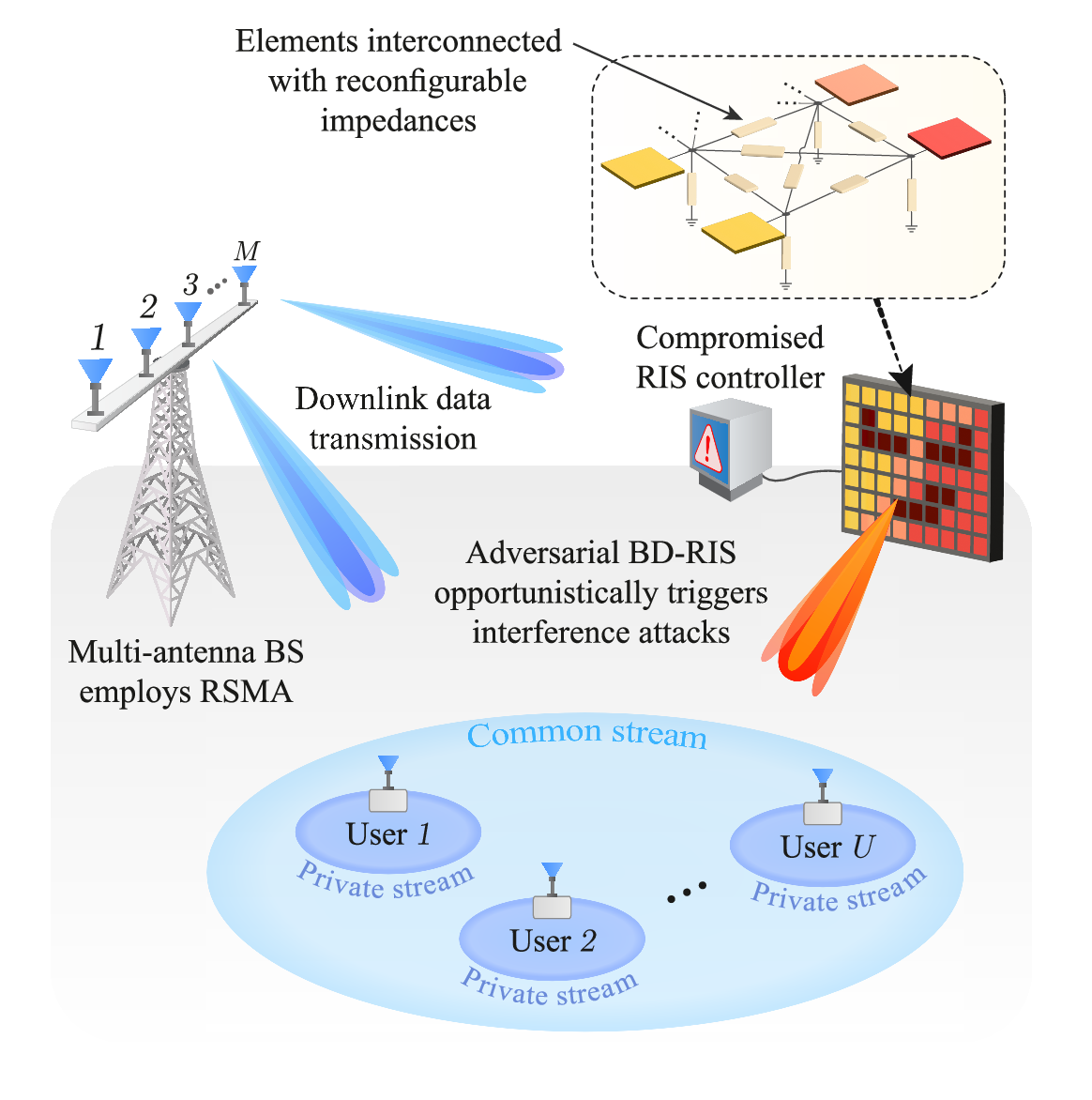}
	\caption{An adversarial entity configures a BD-RIS to launch interference attacks on multiple RSMA users during downlink data transmission.}\label{fig:f1}
\end{figure}

\ac{RSMA} is a split transmission scheme whereby the \ac{BS} encodes a part of the users' data into private messages - transmitted via private precoders -  with the remaining users' data being encoded into a common message - transmitted via a common precoder \cite{Sena22,clerckx2023primer}. On the receiver side, \ac{SIC} is utilized to aid in decoding the individual users' messages. The private precoders are designed as unicast precoders, which - similarly to \ac{SDMA} - makes them sensitive to imperfect \ac{CSI}. The common precoder is constructed as a multicast precoder and, thus, should be more robust to inter-user interference arising due to \ac{CSI} imperfections. Consequently, by adaptively re-allocating the power between the private and common precoders, \ac{RSMA} is able to adapt to \ac{CSI} imperfection~\cite{clerckx2023primer}. In the conference version of this article~\cite{Sena24b}, we observed that \ac{RSMA} is able to adapt also in the face of a channel estimation attack induced by a passive \ac{RIS}, when the \ac{CSI} available at the \ac{BS} is imperfect. However, the extent of \ac{RSMA}'s susceptibility to \ac{RIS}-induced attacks is still an open challenge with many different parts of the \ac{RSMA} protocol, including the uplink pilot training, power allocation, and \ac{SIC} imperfections, requiring deeper investigation. One particularly poignant point is the susceptibility to attacks by novel \ac{RIS} architectures that feature non-diagonal reflection matrices. These architectures, referred to as \acp{BD-RIS}, generalize conventional \ac{RIS} designs (i.e., diagonal \ac{RIS}) by introducing inter-element connections~\cite{li2023reconfigurable}. Such connections enable more advanced optimization capabilities and enhanced energy radiation, which in turn boosts reflection performance at the cost of additional circuit complexity. Nevertheless, the enhanced reflection efficiency of \acp{BD-RIS} may also strengthen their adversarial capabilities, potentially leading to attacks with increased degradation impact. Identifying the distinct vulnerabilities in \ac{RSMA} systems and the advantages that would motivate an attacker to leverage \acp{BD-RIS} constitutes a major research gap.

This article reports on our extensive study and analysis of the robustness of the \ac{RSMA} protocol to attacks against channel acquisition induced by \acp{BD-RIS}, under different attack modes during the uplink pilot training, varying numbers of reflecting elements, and various levels of \ac{CSI} and \ac{SIC} imperfections.
Unlike diagonal \ac{RIS} attacks, where the adversary is limited to independent per-element phase shifts, \ac{BD-RIS}-based attacks exploit inter-element coupling through non-diagonal scattering matrices, causing stronger perturbations to channel estimates than a conventional \ac{RIS} of the same dimension (or, equivalently, achieving similar degradation with fewer reflecting elements).
In our study, we consider three \ac{BD-RIS} element network architectures: \emph{fully connected}, \emph{group-connected}, and \emph{single-connected} (corresponding to the diagonal \ac{RIS}). For each architecture, we propose two different \ac{BD-RIS}-induced interference attacks that exploit the training protocol employed at the \ac{BS}, namely \emph{random} and \emph{aligned}. In both attacks, during uplink pilot training, the \ac{BD-RIS} is configured to absorb or reflect with random reflection coefficients. Then, during data transmission, the \ac{BD-RIS} is configured with newly generated random reflection coefficients (random attack) or with optimized coefficients that maximize the reflected power based on the knowledge of the \ac{RIS} channels (aligned attack). In each case, the proposed algorithm finds a set of reflection coefficients that satisfy the symmetry and unitary constraints of the \ac{BD-RIS}.

Our numerical study reveals that, when perfect \ac{CSI} is available, \ac{RSMA} behaves similarly to \ac{SDMA} and thus is highly susceptible to the attack (corroborating the findings reported in~\cite{huang2023disco,Sena24,Wang24,rivetti2024malicious}), with \ac{BD-RIS} group-connected and fully connected architectures inducing larger performance degradations than a single-connected \ac{RIS}.
However, even though the \ac{BS} is assumed to be completely unaware of the \ac{BD-RIS}-induced manipulation,
\ac{RSMA} displays a surprising degree of robustness when the \ac{CSI} is imperfect, similarly to what we previously reported for conventional \ac{RIS} in \cite{Sena24b}. 
The adaptation mechanism of \ac{RSMA} reacts to the resulting inaccurate \ac{CSI} by allocating more power to the common message and, as a consequence, unintentionally mitigates the impact of the \ac{BD-RIS}-induced attack.
When the transmit power is low (below \unit[10]{dBm}), \ac{RSMA}'s rate is severely degraded by the attack (on par with \ac{SDMA}), but as the transmit power increases, \ac{RSMA} is able to recover from the attack. This trend can be observed across all types of \ac{BD-RIS} attacks. In order to directly quantify this inherent robustness, we propose the \emph{robustness index}, which we define and discuss in subsequent sections. As we gradually adjust the transmit power, the robustness index for \ac{SDMA} under attack steadily decreases towards a saturation point. In contrast, \ac{RSMA}'s robustness index initially declines to a local minimum before beginning a continuous rise, consistently remaining above \ac{SDMA}'s robustness. This highlights the ability of RSMA to sustain performance
under attack, given a sufficiently large transmit power. We also show that the performance degradation induced by group and fully connected architectures becomes increasingly severe compared to a conventional \ac{RIS} as the number of reflecting elements grows\footnote{This work shows that \ac{BD-RIS}-induced attacks can cause significant performance degradation. However, it is noteworthy that practical \ac{BD-RIS} implementations may also be affected by hardware impairments, including phase shift errors (e.g.,~\cite{Li2024RHS_HWI}), mutual coupling (e.g.,~\cite{zheng2024mutual}), and non-ideal response matrix (e.g.,~\cite{de2025performance}). These impairments may also influence the effectiveness of adversarial attacks in practice. A comprehensive investigation of the impact of hardware impairments on adversarial effectiveness is left for future work.}. Our key contributions are as follows:
\begin{itemize}\itemsep1mm
    \item We propose and analyze two distinct attack strategies, employing either random or optimized reflection coefficients, for each \ac{BD-RIS} architecture. These strategies are designed to rigorously satisfy the inherent constraints of the respective \ac{BD-RIS} types, and are targeted at disrupting multi-user data transmission in an \ac{RSMA} system. 
    \item For the random interference attack strategy, we leverage matrix theory to propose a simple algorithm based on Takagi factorization. This algorithm generates valid random \ac{BD-RIS} reflection coefficient matrices that are symmetric and unitary (or block-wise symmetric and unitary for group-connected \ac{RIS} architectures).
    \item For the aligned interference attack strategy, we develop an algorithm to maximize the unwanted reflected power towards legitimate users. To address the resulting complex optimization problem, we propose exploiting the symmetric structure of the \ac{BD-RIS} reflection matrices by using duplication matrices. This approach reformulates the problem into a simplified \ac{QCQP} that can be efficiently solved using \ac{SVD}, followed by projections onto the unitary manifold via the Takagi factorization.
    \item We propose and investigate two metrics to quantify the robustness of \ac{RSMA} systems against \ac{BD-RIS}-induced attacks: \emph{rate degradation} and \emph{robustness index}, ranging from $0$ (minimum robustness) to $1$ (maximum robustness). Our findings demonstrate that under imperfect \ac{CSI}, \ac{RSMA} exhibits substantially greater robustness compared to non-adaptive protocols such as \ac{SDMA}. Moreover, the results highlight the scalability of \ac{RSMA}, as its degradation remains consistently lower than that of \ac{SDMA}, with the gap widening as the \ac{RIS} dimension increases.
    \item Furthermore, we reveal that \ac{BD-RIS} offers a significant adversarial advantage over conventional \ac{RIS} across all investigated scenarios, both under random and aligned interference attacks, an advantage that becomes particularly pronounced when the attacking \ac{RIS} operates in a reflective mode during the uplink pilot training phase. At the same time, we uncover that while \ac{RSMA} is fundamentally more robust to different sources of interference, its robustness strongly depends on the quality of \ac{SIC}. Even small imperfections in \ac{SIC}
    can cause a significant decrease in the system sum rate. 
\end{itemize}

\vspace{1mm}
\noindent  {\it Notation:}
 Boldface lower-case letters denote vectors and upper-case letters represent matrices. The $\ell_2$ norm of a vector $\mathbf{a}$ is denoted by $\|\mathbf{a}\|_2$, and the Frobenius norm of a matrix $\mathbf{A}$ by $\|\mathbf{A}\|_F$. The $i$th column of a matrix $\mathbf{A}$ is denoted by $[\mathbf{A}]_{:,i}$, the transpose and Hermitian transpose of $\mathbf{A}$ are represented by $\mathbf{A}^T$ and $\mathbf{A}^H$, respectively, and $\otimes$ represents the Kronecker product. The operator $\mathrm{vec}(\cdot)$ transforms a matrix into a column vector by stacking its columns sequentially, $\mathrm{unvec}(\cdot)$ reshapes a vectorized matrix into its original dimensions, and $\mathrm{vech}(\cdot)$ is the half-vectorization operator, which stacks the elements from the lower triangular part (including the main diagonal) of a square matrix into a column vector. Moreover, $\mathrm{vecd}(\cdot)$ converts the diagonal elements of a square matrix into a column vector, $\mathrm{diag}(\cdot)$ transforms a vector into a diagonal matrix, and $\mathrm{bdiag}(\cdot)$ constructs a block diagonal matrix sequentially formed by the input matrices.

\section{Related Work}

The concept of an adversary that configures \ac{RIS} to launch attacks against the wireless link was originally proposed in~\cite{lyu2020irs}. The proposed attack relied on using the \ac{RIS} to create cancellation signals. Although the attack has the potential to cause significant power degradation at the receiver, it requires highly accurate \ac{CSI}, which would limit its feasibility. Notably, many works that followed showed that a \ac{RIS}-induced attack can be made effective with limited or even no \ac{CSI}~\cite{staat2022mirror,huang2023disco,huang2024disco,Sena24,rivetti2024malicious}, as long as it takes advantage of the fact that for many wireless protocols the \ac{CSI} acquisition is sufficiently separated in time from data transmission. For instance, in~\cite{staat2022mirror}, it was shown that the channel equalization can be disrupted by the operation of the adversarial \ac{RIS} that randomly flips its reflection pattern between channel acquisition and symbol transmission. This random flipping is particularly harmful to multiple access methods that rely on multi-antenna systems, such as \ac{SDMA}. In this case, adversarial \ac{RIS} increases \ac{CSI} inaccuracy, making linear precoders ineffective in removing inter-user interference. Exactly this kind of attack with randomly set \ac{RIS} coefficients, humorously referred to by their authors as the \yale{disco-ball attack}, was proposed in~\cite{huang2023disco,huang2024disco}. The optimized version of this attack, which aligns the \ac{RIS} channels to boost interference, was proposed and studied in~\cite{Sena24,rivetti2024malicious}. Other protocols and services, such as beam management~\cite{gomes2024beam}, physical layer key generation~\cite{li2024ris-jamming,wang2024beyond}, and integrated sensing and communication~\cite{huang2024integrated,rivetti2025destructive}, have also been shown to be vulnerable to the attack with \ac{RIS} flipping its coefficients between different phases of the wireless protocol. However, many open questions remain in studying \ac{RIS}-induced attacks, particularly as they relate to identifying protocols and methods that may be robust or resilient to the proposed attacks.

\ac{RSMA} was introduced as a multiple access strategy that exploits a split transmission mechanism to adaptively treat inter-user interference at either the \ac{BS} or receiver side, depending on the accuracy of the available \ac{CSI}~\cite{Sena22, clerckx2023primer}. When \ac{CSI} is perfect, \ac{RSMA} resorts to linear precoding at the \ac{BS} that can effectively take care of the inter-user interference. On the other hand, when \ac{CSI} is inaccurate, \ac{RSMA} employs multicast transmission and \ac{SIC} at the receiver side to decode (part of) inter-user interference. There are a variety of benefits that come from this increased flexibility, such as increased spectral and energy efficiency~\cite{zhou2021rate}, improved outage~\cite{lu2024outage}, and reduced latency~\cite{xu2022rate,pala2-24spectral}. \ac{RSMA} has also been shown to be robust and resilient to user mobility~\cite{dizdar2021rate}, binary link failures~\cite{reifert2023comeback}, and, as hinted at earlier, inaccurate \ac{CSI}~\cite{clerckx2023primer}.

The benign application of \acp{RIS} to \ac{RSMA} has been extensively investigated in the literature. In \cite{Sena22}, the interplay between \ac{RSMA} and \ac{RIS} was discussed, highlighting how rate splitting can exploit the additional degrees of freedom introduced by programmable propagation environments. Several works have also considered joint design and optimization of \ac{RSMA} and \acp{RIS}. For instance, \cite{Katwe22} studied uplink \ac{RSMA} in multi-\ac{RIS}-aided millimeter-wave systems with joint optimization of user clustering, power allocation, and active and passive beamforming. The work in \cite{Zhou23STAR} investigated \ac{RSMA} assisted by simultaneous transmitting and reflecting \ac{RIS} over spatially correlated channels. More recently, \cite{pala2-24spectral} considered the joint optimization of transmit precoding, power allocation, and \ac{RIS} coefficients under finite blocklength assumptions, while \cite{ISACRSMA24} considered \ac{RIS}-assisted \ac{RSMA} in integrated sensing and communication networks.
All these works use \ac{RIS} as a performance-enhancing technology and focus on joint transmission design under channel uncertainty models arising from conventional channel estimation and acquisition processes.

In adversarial settings, the RIS acts as an illegitimate component beyond the control of the BS, preventing the joint RIS-RSMA transmission design. It is thus important to consider how RSMA, which explicitly adapts its interference management strategy to channel conditions, behaves when the propagation environment is deliberately manipulated by an illegitimate \ac{RIS}. In a preliminary conference study, we were the first to investigate this question by examining the behavior of RSMA in the presence of an adversarial single-connected RIS~\cite{Sena24b}. There, we observed that RSMA remains surprisingly robust to RIS-induced attacks when already operating with imperfect channel knowledge. This naturally motivates extending the analysis to more general and powerful BD-RIS architectures.

On the \ac{BD-RIS} front, despite being a relatively recent concept, it has attracted significant attention from the research community. The foundational work in \cite{shen2022scattering} modeled \ac{RIS} as a multi-port network using scattering-parameter analysis, enabling the characterization of inter-element coupling and structural constraints beyond diagonal architectures. Building on this perspective, \cite{li2023bd_ris_modes} investigated single-connected, group-connected, and fully connected BD-RIS architectures, while a broader overview of their role in future wireless systems was presented in \cite{li2023reconfigurable}. Reduced-complexity non-diagonal \ac{RIS} designs that approach the performance of fully connected architectures were further studied in \cite{li2022nondiagonal_tvt}, and low-complexity beamforming methods for BD-RIS-aided multi-user networks were proposed in \cite{fang2024lowcomplexity_bd}.
The integration of BD-RIS with \ac{RSMA} has also been investigated. In \cite{fang2022fullyconnected_ris_rsma}, fully connected \ac{RIS} architectures were considered and jointly optimized with \ac{RSMA} transmit beamformers to maximize the downlink sum rate. Robust BD-RIS-assisted \ac{RSMA} designs under imperfect \ac{CSI} were studied in \cite{li2024synergy_bd_ris_rsma}, while \cite{soleymani2024bd_ris_urllc} investigated BD-RIS-assisted \ac{RSMA} under finite blocklength constraints for \ac{URLLC} scenarios. Uplink \ac{RSMA} with BD-RIS was recently considered in \cite{khisa2025meta_uplink} using learning-based optimization, and BD-RIS-assisted \ac{RSMA} has also been studied in the context of simultaneous wireless information and power transfer in \cite{asif2025rsma_swipt_bdris}.

\ac{BD-RIS} architectures have also been investigated from an adversarial and security-oriented perspective. In \cite{wang2024non-diagonal} and \cite{wang2024beyond}, non-diagonal \ac{RIS} were shown to enable channel reciprocity attacks and to disrupt physical-layer key generation without active jamming. Passive omnidirectional jamming using \yale{DISCO surfaces} was proposed in \cite{huang2024disco-omni}. More recently, \cite{Iivanainen25} analyzed worst-case adversarial channels induced by BD-RIS and identified fundamental limits of simplified surface architectures. 
Although these works establish the adversarial potential of \acp{BD-RIS}, the impact of such attacks on next-generation multiple access schemes, particularly on the robustness of \ac{RSMA}, remains largely unexplored.

In the conference version of this work, we took a first step toward addressing this gap by studying the impact of adversarial diagonal \ac{RIS} on \ac{RSMA}~\cite{Sena24b}. 
In this article, we extend our preliminary study by quantifying the robustness of \ac{RSMA} to adversarial \ac{BD-RIS} attacks. The details of the \ac{RSMA} system model, BD-RIS architectures, and the considered attack mechanisms are introduced next.

\section{System Model}

Consider the multi-user \ac{MISO} system depicted in \Fig{f1}, where a \ac{BS} equipped with a linear array of $M$ antennas communicates with $U$ single-antenna users, represented by the index set~$\mathcal{U} = \{1,2,\ldots, U\}$, with the aid of \ac{RSMA}. The system operates in the presence of an adversarial entity that controls a \ac{BD-RIS} comprising $D$ reflecting elements. The attacker manipulates the \ac{BD-RIS} to disrupt the channel estimation phase at the \ac{BS}, configuring it to either absorb incoming pilots \cite{Sena24, huang2023disco} (absorption mode) or reflect them randomly \cite{huang2024disco} (reflective mode). The impact of the mode choice will be investigated in Section~\ref{res_sec}. 

\begin{figure*}[t]
   \centering
    \includegraphics[width=.8\textwidth]{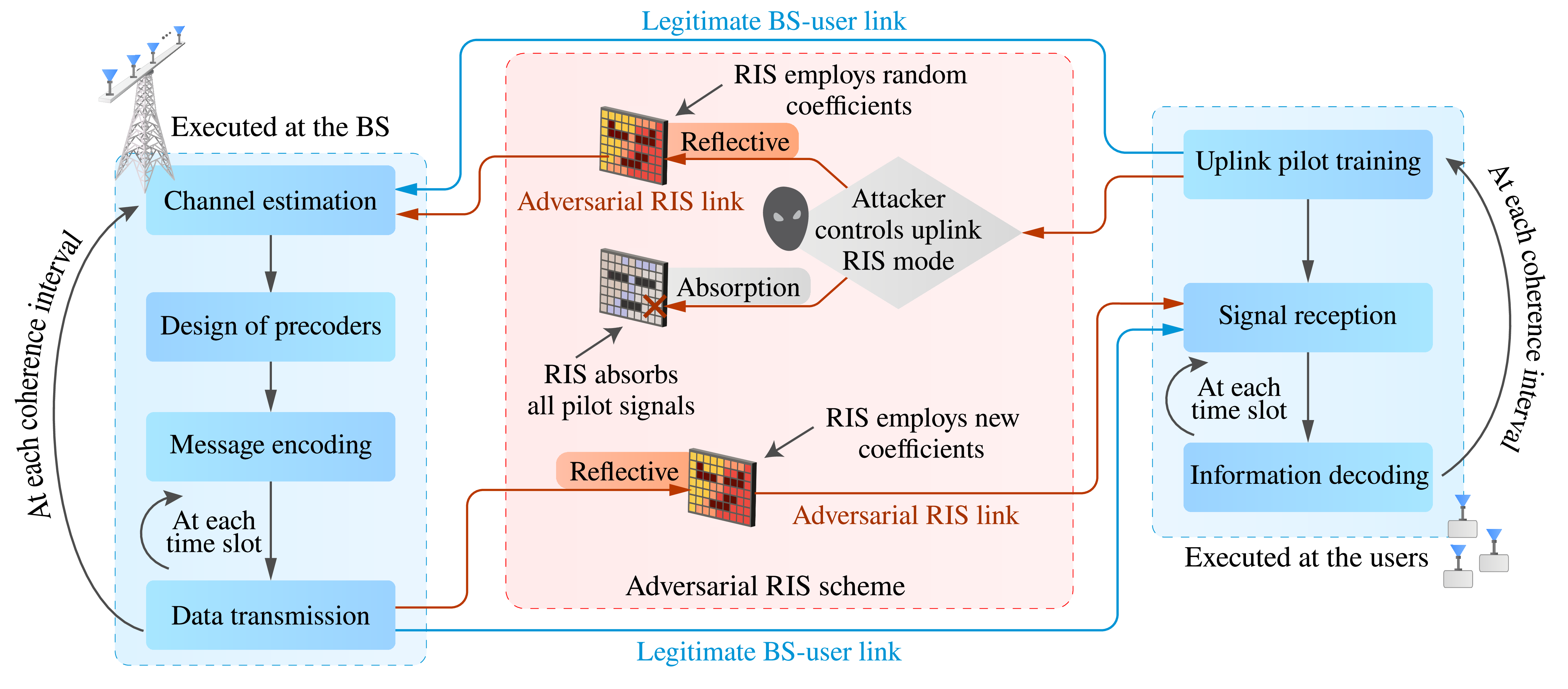}
   \caption{Simplified diagram of the RIS-induced attack strategy.}
  \label{adv_protocol}
\end{figure*}

Subsequently, the attacker generates a new set of reflecting coefficients during the data transmission phase, altering the propagation environment in a manner unknown to the \ac{BS}. This ensures that the \ac{BS} operates with corrupted \ac{CSI}, as its estimates from the training phase do not reflect the RIS's true configuration during data transmission. The discrepancy between estimated and actual \ac{CSI} reduces the performance of precoders, ultimately leading to inter-user interference. Fig.~\ref{adv_protocol} illustrates the overall adversarial protocol.

Note that the attacker must ensure that its RIS configurations differ between the uplink training and downlink data transmission phases. As explained in \cite{huang2024disco}, this can be accomplished by generating a new set of coefficients on a timescale comparable to or shorter than the pilot training duration. A more sophisticated alternative is for the adversary to actively synchronize its attack. Since pilot sequences and their transmission intervals are typically public information defined by the wireless standard (e.g., in LTE Advanced and 5G \ac{NR} systems), an attacker can passively monitor the channel to detect these periodic transmissions and determine the timing of the uplink training phase \cite{Akgun19}.

\begin{figure}[t]
	\centering
	\includegraphics[width=1.0\linewidth]{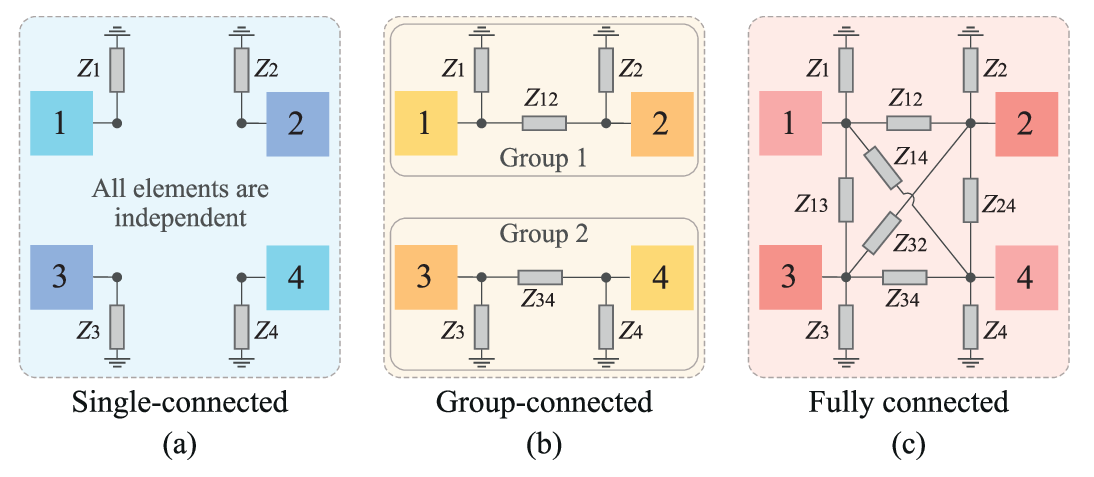}
	\caption{Illustrative comparison of different \ac{RIS} architectures with four elements: (a) single-connected; (b) group-connected; and (c) fully connected. $Z_i$ represents the reconfigurable self-impedance of element $i$, while $Z_{ij}$ denotes the mutual impedance between elements $i$ and $j$, with $i, j \in \{1, \ldots, 4\}$.}
	\label{fig:ris_architectures}
\end{figure}

\subsection{BD-RIS Architectures}
We consider two \ac{BD-RIS} architectures: fully connected and group-connected.
Fig.~\ref{fig:ris_architectures} illustrates the structural differences between these \ac{BD-RIS} architectures and the conventional (single-connected) \ac{RIS}.
Specifically, in the fully connected \ac{RIS}, all elements are interconnected through reconfigurable impedances.
A fully connected \ac{RIS} with $D$ reflecting elements operates as a $D$-port reciprocal network. The interaction among elements is captured by the scattering matrix $\bm{\Theta} \in \mathbb{C}^{D \times D}$, which characterizes the reflection behavior of the system. Energy conservation imposes the constraint $\bm{\Theta} \bm{\Theta}^H \preceq \mathbf{I}_{D}$, implying that the Frobenius norm satisfies $\frac{1}{\sqrt{D}} |\bm{\Theta}|_F \leq 1$. When the circuit is lossless, the scattering matrix becomes unitary, i.e., $\bm{\Theta} \bm{\Theta}^H = \mathbf{I}_{D}$. Furthermore, due to the reciprocal nature of the \ac{BD-RIS} hardware, the symmetry condition $\bm{\Theta} = \bm{\Theta}^T$ must also be satisfied\footnote{The concept of non-reciprocal BD-RIS, featuring non-symmetric unitary reflection matrices, has also been proposed recently \cite{Li2025NonReciprocalBDRIS}. Such architectures can break channel reciprocity and induce a mismatch between uplink and downlink cascaded channels even if the RIS configuration is kept fixed over time, thereby facilitating adversarial attacks.}.
This property implies that $\frac{D(D+1)}{2}$ independent coefficients are needed to fully define the entries of $\bm{\Theta}$.

The group-connected architecture introduces a balance between complexity and performance. In this setup, the reflecting elements are divided into $G$ independent groups, each of which forms a fully connected subnetwork. As a result, the overall scattering matrix exhibits a block diagonal structure:
\begin{align}\label{gc_ris_mat}
\bm{\Theta} = \mathrm{bdiag}(\bm{\Theta}_1, \cdots, \bm{\Theta}_G) \in \mathbb{C}^{D\times D},
\end{align}
where each submatrix $\bm{\Theta}_g \in \mathbb{C}^{D_g \times D_g}$ models the signal reflections induced by its corresponding fully connected group, with $D_g$ denoting the number of elements of the $g$th group, such that $D = \sum_g^G D_g$. Similar to the fully connected case, each submatrix satisfies the conditions $\bm{\Theta}_g = \bm{\Theta}_g^T$ and $\bm{\Theta}_g \bm{\Theta}_g^H \preceq \mathbf{I}_{D_g}$. The number of independent parameters needed to configure each group is, therefore, $\frac{D_g(D_g+1)}{2}$, in which $D_g \leq D$. From this point onward, we consider ideal reflection models, i.e., we assume $\bm{\Theta} \bm{\Theta}^H = \mathbf{I}_{D}$ and $\bm{\Theta}_g \bm{\Theta}_g^H = \mathbf{I}_{D_g}$ for the fully and group-connected RIS, respectively. The impact of different adversarial strategies is analyzed in the following sections.

\subsection{RSMA Protocol}

In RSMA, the \ac{BS} partitions each user's message into common and private parts. Then, all common parts are encoded into a single symbol $s^c$, while each private part, intended for a specific user $u$, is modulated into a private symbol $s^p_u$.

Next, based on the acquired \ac{CSI}, the \ac{BS} designs two linear precoders: $\mathbf{w}^{c} \in \mathbb{C}^{M}$ to deliver the common symbol $s^c$ to all users and $\mathbf{w}^{p}_{u} \in \mathbb{C}^{M}$ to transmit the private symbol $s^p_u$ to its intended user. Finally, the precoded symbols are superimposed in the power domain, resulting in the transmitted signal:
\begin{align}\label{trans_sig}
    \mathbf{s} &= \mathbf{w}^c  \sqrt{P\alpha^c} s^c + \sum_{i \in \mathcal{U}} \mathbf{w}^{p}_{i} \sqrt{P\alpha_{i}^p} s_{u}^p  \in \mathbb{C}^{M},
\end{align}
where $\alpha^c$ and $\alpha_{i}^p$ are the power coefficients for the common and private symbols, respectively, satisfying $\alpha^c + \sum_{i\in \mathcal{U}} \alpha_{i}^p = 1$, with $P$ denoting the total transmission power. Users receive the superimposed data streams through both the direct link and the adversarial \ac{BD-RIS} reflections. Thus, the received signal at user $u$ is given by:
\begin{align}\label{rec_sig_01}
    r_{u} = \mathbf{g}_{u}^H \bm{\Theta} \mathbf{G}\mathbf{s} + \mathbf{h}_{u}^H \mathbf{s} + n_{u},  
\end{align}
where $\mathbf{h}_{u} \in \mathbb{C}^{M}$, $\mathbf{G} \in \mathbb{C}^{D \times M}$, and $\mathbf{g}_{u} \in \mathbb{C}^{D}$ model the propagation channels for the direct \ac{BS}-user link, the \ac{BS}-\ac{RIS} link, and the \ac{RIS}-user link, respectively, and $n_{u} \in \mathbb{C}$ denotes the additive white Gaussian noise with variance $\sigma^2$.  

Each user first decodes the common message, while treating all private signals as interference. Consequently, the \ac{SINR} for the common message at user $u$ is:
\begin{align}\label{common_sinr}
    \gamma^c_u = \frac{|\big(\mathbf{g}_{u}^H\bm{\Theta}\mathbf{G} +  \mathbf{h}_{u}^H\big) \mathbf{w}^c|^2P \alpha^c}{\sum_{i \in \mathcal{U}} | \big(\mathbf{g}_{u}^H\bm{\Theta}\mathbf{G} +  \mathbf{h}_{u}^H\big) \mathbf{w}^{p}_{i}|^2 P\alpha_{i}^p + \sigma^2}.
\end{align}
Since all users must successfully decode the common message, the corresponding achievable rate is dictated by the weakest user, resulting in:
\begin{align}\label{eq:common_rate}
    R^c = \min_{u \in \mathcal{U}} R_u^c, \quad \text{where} \quad R_u^c = \log_2(1 + \gamma_u^c).
\end{align}

Once the common message is recovered, \ac{SIC} is applied to remove its contribution from $r_u$. 
However, due to factors such as hardware impairments and estimation errors, the user's reconstructed common signal might not perfectly cancel the received one. This mismatch leads to imperfect \ac{SIC}, which can be modeled as a linear function of the common signal power\cite{Sena20}:
\begin{align} \label{eq:sic_error}
    \chi^c_u = \xi \left| (\mathbf{g}_{u}^H\bm{\Theta}\mathbf{G} + \mathbf{h}_{u}^H) \mathbf{w}^c \right|^2 P \alpha^c,
\end{align}
where $\xi \in [0,1]$ denotes the \ac{SIC} error factor. A smaller $\xi$ indicates a higher quality of \ac{SIC}, with $\xi=0$ representing the ideal case of perfect cancellation. This residual interference is treated as additional noise when decoding the private message. Consequently, the private \ac{SINR} at user $u$ is given by:
\begin{align}\label{private_sinr}
    \gamma^p_u = \frac{|\big(\mathbf{g}_{u}^H\bm{\Theta}\mathbf{G} +  \mathbf{h}_{u}^H\big) \mathbf{w}^{p}_{u}|^2 P \alpha_{u}^p}{\sum_{i \in \mathcal{U}, i \neq u} | \big(\mathbf{g}_{u}^H\bm{\Theta}\mathbf{G} +  \mathbf{h}_{u}^H\big) \mathbf{w}^{p}_{i}|^2 P \alpha_{i}^p + \chi^c_u + \sigma^2}.
\end{align}

The achievable private rate at user $u$ is then:
\begin{align}\label{eq:private_rate}
    R_u^p = \log_2 (1 + \gamma_u^p),
\end{align}
yielding a system sum rate of: 
\begin{align}\label{sum_rate}
   R = R^c + \sum_{i \in \mathcal{U}} R_i^p.
\end{align}

\subsection{Quantifying Robustness of RSMA}

We propose two performance metrics to evaluate the robustness of \ac{RSMA} against \ac{RIS}-induced attacks, namely \emph{rate degradation} and \emph{robustness index}. The rate degradation quantifies the performance loss caused by an attack, defined as the difference between the baseline performance in a secure environment and the degraded performance during an attack. To this end, the common rate is assumed to be uniformly allocated among the users, i.e., $R^c/U$. The overall effective rate for user $u$ is then the sum of this portion of the common rate and the individual private rate $R^p_u$. The rate degradation experienced by the $u$th user is defined as
\begin{align}\label{degradation}
    \Delta R_{u} = \bar{R}^p_u - \tilde{R}^p_u + \frac{\bar{R}^c - \tilde{R}^c}{U},
\end{align}
where $\tilde{R}^p_u$ and $\tilde{R}^c$ are the private and common rates observed when the system is under attack, and $\bar{R}^p_u$ and $\bar{R}^c$ are the reference rates in the absence of the attack.

The robustness index $\kappa$ complements the rate degradation metric by quantifying the proportion of baseline performance retained during an attack. This normalized metric quantifies the system's ability to maintain performance under \ac{RIS}-induced attacks relative to its baseline rates $\bar{R}^p_u$ and $\bar{R}^c$. The robustness index  is defined as  
\begin{align}\label{robustness}
    \kappa = \frac{1}{U} \sum_{u=1}^{U} \left( 1 -  \frac{\Delta R_{u}}{\frac{\bar{R}^c}{U} + \bar{R}^p_u} \right). 
\end{align}  
Note that a higher degradation $\Delta R_{u}$ leads to a lower robustness index $\kappa \in [0,1]$, such that $\kappa = 0$ indicates complete collapse, while $\kappa = 1$ corresponds to complete robustness.

\subsection{CSI Acquisition}
Under the assumption of reciprocity, the downlink channels can be estimated at the \ac{BS} based on uplink measurements from pilot signals transmitted by the users. As explained, during this phase, the attacker can configure the \ac{BD-RIS} to either absorb or randomly reflect incoming signals. In absorption mode, the pilot signals impinging on the \ac{BD-RIS} are prevented from reaching the \ac{BS}. In the random reflective mode, on the other hand, the pilot signals will characterize the \ac{BD-RIS}-related propagation paths. However, the attacker reconfigures the reflection coefficients during the data transmission phase, rendering the acquired \ac{CSI} inaccurate. Thus, in both modes, only imperfect estimates of the channels $\mathbf{h}_u$ are accessible to the \ac{BS}. Adding to the \ac{BD-RIS} distortions, other rapid changes in the propagation environment and hardware limitations can also introduce errors in the estimation process. As a result, the direct link estimates will be a combination of the true channels and error components, modeled as:
\begin{align}\label{eq:imp_csi_BU}
  \hat{\mathbf{h}}_u = \sqrt{1-\epsilon^{\text{\tiny BS-U}}}\, (\mathbf{h}_u + \psi\mathbf{G}^H\bm{\Theta}\mathbf{g}_u) + \sqrt{\epsilon^{\text{\tiny BS-U}}}\, \mathbf{e}^{\text{\tiny BS-U}}_u,
\end{align}
where $\hat{\mathbf{h}}_u \in \mathbb{C}^{M}$ is the estimated channel, $\mathbf{e}^{\text{\tiny BS-U}}_u \in \mathbb{C}^{M}$ is a standard complex Gaussian error vector, $\psi$ is a binary coefficient modeling the operation mode of the \ac{BD-RIS}, with $\psi = 0$ corresponding to absorption mode, and $\psi = 1$ corresponding to reflective mode, and $\epsilon^{\text{\tiny BS-U}} \in [0,1]$ is the error factor for the BS-user link.
This model indicates that when $\epsilon^{\text{\tiny BS-U}}=0$, the channel estimate is perfect (i.e., $\hat{\mathbf{h}}_u = \mathbf{h}_u + \psi\mathbf{G}^H\bm{\Theta}\mathbf{g}_u$), while increasing values of $\epsilon^{\text{\tiny BS-U}}$ imply a larger error component in the estimate. Similarly, when $\psi = 1$, the uplink \ac{CSI} is further perturbed by the adversarial \ac{BD-RIS} link.

As for the attacker, we assume it can only estimate the channels associated with its \ac{BD-RIS} and that these estimates are imperfect. Specifically, the estimates of the BS-RIS channel matrix and the RIS-user channel vector are given by
\begin{align}\label{eq:imp_csi_BR}
  \hat{\mathbf{G}} = \sqrt{1-\epsilon^{\text{\tiny BS-RIS}}}\, \mathbf{G} + \sqrt{\epsilon^{\text{\tiny BS-RIS}}}\, \mathbf{E}^{\text{\tiny BS-RIS}},
\end{align}
and
\begin{align}\label{eq:imp_csi_RU}
  \hat{\mathbf{g}}_u = \sqrt{1-\epsilon^{\text{\tiny RIS-U}}}\, \mathbf{g}_u + \sqrt{\epsilon^{\text{\tiny RIS-U}}}\, \mathbf{e}^{\text{\tiny RIS-U}}_u,
\end{align}
where $\mathbf{E}^{\text{\tiny BS-RIS}}$ and $\mathbf{e}^{\text{\tiny RIS-U}}_u$ are the corresponding complex Gaussian error components, with $\epsilon^{\text{\tiny BS-RIS}}$ and $\epsilon^{\text{\tiny RIS-U}}$ denoting the error factors for the BS-RIS and RIS-user links, respectively.

It is noteworthy that channel acquisition is a well-known practical challenge in \ac{RIS}-aided systems, due to limited sensing hardware, calibration and synchronization challenges, or insufficient observation time due to mobility, and even more so when the \ac{RIS} operates in a stand-alone manner without direct coordination with the \ac{BS}, as would be the case for an attacker. One potential solution is to exploit architectures with embedded sensing capabilities, such as the concept of a semi-passive \ac{RIS} proposed in \cite{Jin21}, where a small subset of elements is equipped with dedicated sensing circuitry to acquire partial \ac{CSI} and reconstruct the remaining channels at the \ac{RIS} controller by passively sensing existing pilot transmissions. A related approach is the hybrid reflecting-and-sensing \ac{RIS} \cite{Zhang2021SRSHRIS}, in which elements are dual-purpose, simultaneously reflecting and sensing a portion of the impinging signal to enable channel estimation from existing pilots. In a realistic adversarial deployment, however, such acquisition remains challenging, resulting in imperfect estimates, as we assume in our model.

\subsection{BS Precoder Design}
The private precoder should ensure that the interference observed at user $u$ due to transmissions intended for any other user $u'$ is negligible, i.e., $\mathbf{h}_{u'}^H \mathbf{w}_u^p \approx 0$, $\forall \, u' \neq u$. This objective is achieved via a regularized zero-forcing approach\footnote{Other precoding strategies could also be considered. For instance, iterative \ac{RSMA} precoding strategies that jointly optimize transmit precoders and power allocation are expected to achieve higher absolute sum rates, e.g., \ac{WMMSE}-based designs \cite{Joudeh2016}, at the expense of a greater computational complexity. Incorporating WMMSE and other strategies into the proposed framework and investigating the impact on \ac{RSMA} robustness is an interesting direction for future work.}, which effectively mitigates inter-user interference while balancing noise enhancement. Specifically, the BS first constructs the estimated direct-link channel matrix as
\begin{align}
  \hat{\mathbf{H}} = \begin{bmatrix} \hat{\mathbf{h}}_1, \hat{\mathbf{h}}_2, \cdots, \hat{\mathbf{h}}_U \end{bmatrix} \in \mathbb{C}^{M \times U}.
\end{align}
Then, assuming that $M \geq U$, which guarantees that $\hat{\mathbf{H}}$ has full column rank, the regularized zero-forcing precoding matrix can be computed as
\begin{align}
  \mathbf{W}^p = \hat{\mathbf{H}} \left( \hat{\mathbf{H}}^H \hat{\mathbf{H}} + \omega \mathbf{I}\right)^{-1} \in \mathbb{C}^{M \times U},
\end{align}
where the regularization factor is given by $\omega = {\sigma^2}/{P}$. The private precoder for user $u$ can be readily obtained by
\begin{align}\label{eq:private_prec}
  \mathbf{w}_u^p = \beta^p_u {\left[ \mathbf{W}^p \right]_{:,u}},
\end{align}
where $\beta^p_u = \frac{1}{ (\mathbf{w}_u^p)^H\mathbf{w}_u^p}$ is the normalization factor for user $u$.

The common precoder $\mathbf{w}^{c}$ should reliably deliver the common message across all users. However, its optimal design is NP-hard, while approximate solutions typically require iterative methods with high computational complexity~\cite{Konar17}. To tackle this challenge, we adopt a low-complexity weighted \ac{MF} approach as follows:
\begin{align}\label{eq:common_precoder}
    \mathbf{w}^{c} = \beta^c \sum_{i\in \mathcal{U}} \upsilon_i \hat{\mathbf{h}}_{i},
\end{align}
where the normalization factor $\beta^c = \left\|\sum_{i\in \mathcal{U}} \upsilon_i \hat{\mathbf{h}}_{i}\right\|_2^{-1}$ guarantees that $\|\mathbf{w}^{c}\|_2 = 1$, and $\upsilon_i$ is the weight for the $i$th user. In particular, we prioritize users with weaker channel conditions by setting $\upsilon_i = 1/{\|\hat{\mathbf{h}}_{i}\|_2^2}$. This strategy ensures that users experiencing weaker channels receive higher priority in the common precoder design.

\subsection{Power Allocation}\label{pa_subsection}

The BS employs a power allocation policy aiming at the maximization of the system sum rate. Given the precoders $\mathbf{w}^c$ and $\mathbf{w}^p_u$ for each user $u \in \mathcal{U}$, the BS aims to solve the following optimization problem:
\begin{subequations}\label{power_prob}
\begin{align}
    \max_{\alpha^c,\, \{\alpha_u^p\}_{u\in\mathcal{U}}} \quad & R^c + \sum_{u\in\mathcal{U}} R_u^p, \\
    \text{s.t.} \hspace{9mm} & \alpha^c + \sum_{u\in\mathcal{U}} \alpha_u^p = 1, \\
    & \alpha_u^p \ge 0, \quad \alpha^c \ge 0.
\end{align}
\end{subequations}
A closed-form optimal solution is not possible due to the minimum operator in the common rate in \eqref{eq:common_rate}, i.e., $R^c = \min_{u \in \mathcal{U}} R_u^c$, and the coupled power coefficients, which make the problem non-convex. Instead, we propose a simple but effective approach to approximate our objective.

A meaningful way to perform power allocation is to keep the interference resulting from imperfect \ac{CSI} approximately at the same level as the noise floor \cite{clerckx2023primer}. By doing this, the degradation of users' data rates can be mitigated.
In our system, which considers imperfect \ac{SIC}, the noise floor for the private messages is a linear combination of the thermal noise $\sigma^2$ and the residual \ac{SIC} interference $\chi^c_u$. Therefore, we aim to ensure that the SINR expression for the private stream, $\gamma_u^p$, is on the same order as the sum of this effective noise power, i.e., $\sum_{i\neq u} | \mathbf{h}_{u}^H \mathbf{w}^{p}_{i}|^2 P \alpha^p_u \approx \xi \left| \mathbf{h}_{u}^H \mathbf{w}^c \right|^2 P \alpha^c + \sigma^2$.
To achieve this goal, we first adopt a uniform power allocation across the private messages\footnote{Allowing a more general joint power and precoder optimization across private and common streams is expected to achieve higher performance. However, due to the inherent complexity of such a joint design, it is left for future work.}, i.e., $\alpha^{p} = \alpha^{p}_{u}, \forall u\in \mathcal{U}$. Then, we can substitute $\alpha^c = 1 - U\alpha^p$ and show for the $u$th user that:
\begin{align}
    &\sum_{i\neq u} | \mathbf{h}_{u}^H \mathbf{w}^{p}_{i}|^2 P \alpha^p \approx \xi \left| \mathbf{h}_{u}^H \mathbf{w}^c \right|^2 P \alpha^c + \sigma^2 \nonumber\\
    &\implies \alpha^p \propto \frac{\xi |\mathbf{h}_{u}^H \mathbf{w}^c|^2 P + \sigma^2}{\sum_{i\neq u} | \mathbf{h}_{u}^H \mathbf{w}^{p}_{i}|^2 P + \xi U |\mathbf{h}_{u}^H \mathbf{w}^c|^2 P}.\label{power_p_prop}
\end{align}
That is, the desired $\alpha^{p}$ should be directly proportional to the effective noise floor and inversely proportional to the private inter-user interference.

Note that we need to compute an $\alpha^{p}$ that balances the interference terms across all users so as to maximize the system's sum rate. Given this goal, we prioritize the user with the best channel conditions, i.e., the one that contributes the most to the sum rate. To this end, we select as a reference the best performance indicator given by:
\begin{align}
    u^\star = \arg\max_{u\in\mathcal{U}} \hspace{2mm} \frac{| \mathbf{h}_{u}^H \mathbf{w}^{p}_{u}|^2 P}{\sum_{i\neq u} | \mathbf{h}_{u}^H \mathbf{w}^{p}_{i}|^2 P + \xi |\mathbf{h}_{u}^H \mathbf{w}^c|^2P} + \sigma^2.
\end{align}

After determining the reference user, we recall \eqref{power_p_prop} and transform the allocation into a tractable problem by introducing a scaling factor $\eta$, as follows:
\begin{align}\label{private_policy}
    \alpha^p = \frac{\eta \left( \xi |\mathbf{h}_{u^\star}^H \mathbf{w}^c |^2 P + \sigma^2 \right)}{\sum_{i\neq u^\star} | \mathbf{h}_{u^\star}^H \mathbf{w}^{p}_{i}|^2 P + \xi U |\mathbf{h}_{u^\star}^H \mathbf{w}^c |^2 P}.
\end{align}
The common power coefficient is consequently given by $\alpha^c = 1 - U \alpha^p$. The scaling factor $\eta$ adjusts the aggressiveness of the allocation, i.e., a higher $\eta$ increases private power allocation, while a lower $\eta$ reserves more power for the common message.

Thus, the original formulation can be transformed into a simpler version with a single optimization variable, $\eta$, resulting in the following relaxed power allocation problem:
\begin{subequations}\label{power_prob_line_search}
\begin{align}
    \max_{\eta} \quad & R^c(\eta) + \sum_{u\in\mathcal{U}} R_u^p(\eta),\\[1mm]
    \text{s.t.} \hspace{4mm} & \alpha^p = \frac{\eta \left( \xi |\mathbf{h}_{u^\star}^H \mathbf{w}^c |^2 P + \sigma^2 \right)}{\sum_{i\neq u^\star} | \mathbf{h}_{u^\star}^H \mathbf{w}^{p}_{i}|^2 P + \xi U |\mathbf{h}_{u^\star}^H \mathbf{w}^c |^2 P},\\[1mm]
    & \alpha^c = 1 - U \alpha^p,\\[1mm]
    & U \alpha^p \ge 0, \quad \alpha^c \ge 0. \label{coeffs_constraint}
\end{align}
\end{subequations}
where $R^c(\eta)$ and $R_u^p(\eta)$ denote the common and private rates as functions of $\eta$. Although the problem remains non-convex due to the minimum operator in the common rate and the nonlinear dependence of the SINRs on $\eta$, a linear search over $\eta$ within its feasible range provides a practical and efficient method to approximate the solution of \eqref{power_prob_line_search}. Specifically, if $\eta$ is chosen such that $0 \le \eta \le \frac{\sum_{i\neq u^\star} | \mathbf{h}_{u^\star}^H \mathbf{w}^{p}_{i}|^2 P + \xi U |\mathbf{h}_{u^\star}^H \mathbf{w}^c |^2 P}{U \left( \xi |\mathbf{h}_{u^\star}^H \mathbf{w}^c |^2 P + \sigma^2 \right)}$, the constraints in \eqref{coeffs_constraint} are automatically satisfied.

This power allocation policy also provides intuition for the robustness trends observed in Section~\ref{res_sec}. Since the \ac{BS} is unaware of any \ac{BD-RIS} manipulation, it performs precoding using the available corrupted \ac{CSI}. The attack is then reflected in the effective \acp{SINR} through increased residual inter-user interference due to beamforming mismatch, i.e., larger leakage terms $\sum_{i\neq u}\!\left|\mathbf{h}_u^H\mathbf{w}_i^p\right|^2$ in the denominator of the private \ac{SINR} in \eqref{private_sinr}. In particular, the proposed policy explicitly ties the private-stream power coefficient to the ratio in \eqref{private_policy}, where the denominator includes the private-stream leakage power $\sum_{i\neq u}\!\left|\mathbf{h}_u^H\mathbf{w}_i^p\right|^2P$. Thus, when the attack increases the leakage term, the resulting $\alpha_p$ decreases and the remaining power is shifted to the common stream, i.e., $\alpha_c=1-U\alpha_p$.
As the transmit power increases, this mechanism becomes more pronounced because the interference terms dominate over $\sigma^2$, which promotes the allocation of more power to the common message. Therefore, even without attack awareness or explicit countermeasures, the \ac{RSMA} adaptation mechanism intended to cope with standard imperfect \ac{CSI} also mitigates the \ac{BD-RIS}-induced degradation.

\section{Design of BD-RIS-Induced Attacks}

The malicious goal of the attacker is to compute a reflection matrix $\bm{\Theta}$ that degrades the performance of the employed \ac{RSMA} scheme. In the following subsections, we investigate two approaches to accomplish the goal.

\subsection{Random Interference Attack}\label{subsec:random_interf}

A simple yet effective method to degrade the performance of \ac{RSMA} involves randomly configuring the reflection coefficients of an adversarial \ac{RIS} to introduce interference, as demonstrated in our previous work with diagonal \acp{RIS} \cite{Sena24}.

\paragraph{Fully Connected Architecture}
As explained, a fully connected \ac{RIS} requires its reflection matrix to be both symmetric and unitary. To obtain a valid set of reflection coefficients, the adversary starts by creating a random complex matrix $\tilde{\bm{\Gamma}} \in \mathbb{C}^{D \times D}$ and then symmetrizes it, as follows:
\begin{align}\label{eq:symmetrization}
\bm{\Gamma} &= \frac{1}{2}\left(\tilde{\bm{\Gamma}} + \tilde{\bm{\Gamma}}^T\right).
\end{align}
However, $\bm{\Gamma}$ is generally not unitary. Therefore, we need to project it onto the set of symmetric unitary matrices while respecting its symmetry constraint. To this end, we can rely on the Takagi factorization, which is introduced next.

\property\label{prop:takagi} 
Let $\bm{\Gamma}$ be a symmetric complex matrix. Then, the Takagi factorization states that there exists a unitary matrix $\mathbf{U}_{\bm{\Gamma}}$ and a diagonal matrix $\bm{\Sigma}_{\bm{\Gamma}}$ such that $\bm{\Gamma} = \mathbf{U}_{\bm{\Gamma}} \bm{\Sigma}_{\bm{\Gamma}} \mathbf{U}_{\bm{\Gamma}}^T$, where the diagonal entries of $\bm{\Sigma}_{\bm{\Gamma}}$ are nonnegative singular values of $\bm{\Gamma}$, called Takagi values, and the columns of $\mathbf{U}_{\bm{\Gamma}}$ are the associated orthonormal Takagi vectors.
With Property~\ref{prop:takagi}, the desired symmetric unitary reflection matrix can be achieved through the following lemma.

\lemma\label{lemma:uni_proj}
Let $\bm{\Gamma}$ be a complex symmetric matrix that admits the Takagi factorization $\bm{\Gamma} = \mathbf{U}_{\bm{\Gamma}} \mathbf{\Sigma}_{\bm{\Gamma}} \mathbf{U}_{\bm{\Gamma}}^T$. Then, the unique symmetric unitary matrix
\begin{align}
\mathbf{\Theta} = \mathbf{U}_{\bm{\Gamma}}\mathbf{U}_{\bm{\Gamma}}^T,
\end{align}
minimizes the Frobenius norm $\|\bm{\Gamma} - \mathbf{\Theta}\|_F$ and, thus, defines the optimal projection onto the set of symmetric unitary matrices.

\textit{Proof:}  
Since $\bm{\Gamma}$ is complex symmetric, its Takagi factorization is given by $\bm{\Gamma} = \mathbf{U}_{\bm{\Gamma}} \mathbf{\Sigma}_{\bm{\Gamma}} \mathbf{U}_{\bm{\Gamma}}^T$. Any symmetric unitary matrix can be written in the form $\mathbf{\Theta} = \mathbf{U}_{\bm{\Gamma}}\mathbf{D}\mathbf{U}_{\bm{\Gamma}}^T$, where $\mathbf{D}$ is a diagonal matrix with entries on the unit circle, i.e., $|[\mathbf{D}]_{ii}| = 1$, for each diagonal element. The problem of finding the symmetric unitary matrix closest to $\bm{\Gamma}$ reduces to choosing $\mathbf{D}$ such that
\begin{align}
\|\bm{\Gamma} - \mathbf{\Theta}\|_F &= \|\mathbf{U}_{\bm{\Gamma}} \mathbf{\Sigma}_{\bm{\Gamma}} \mathbf{U}_{\bm{\Gamma}}^T - \mathbf{U}_{\bm{\Gamma}}\mathbf{D}\mathbf{U}_{\bm{\Gamma}}^T\|_F \nonumber\\
&= \|\mathbf{\Sigma}_{\bm{\Gamma}} - \mathbf{D}\|_F
\end{align}
is minimized, where the last equality follows from the unitary invariance of the Frobenius norm. Since each diagonal entry $[\bm{\Sigma}_{\bm{\Gamma}}]_{ii}$ is nonnegative, for each $i$ the distance 
$\left|[\bm{\Sigma}_{\bm{\Gamma}}]_{ii} - [\mathbf{D}]_{ii} \right|$ is minimized by choosing 
$[\mathbf{D}]_{ii} = 1$, as $\left|[\bm{\Sigma}_{\bm{\Gamma}}]_{ii} - e^{i\theta} \right|$ attains its 
minimum when $\theta = 0$. Therefore, the optimal projection is given by $\mathbf{D} = \mathbf{I}$, yielding $\mathbf{\Theta} = \mathbf{U}_{\bm{\Gamma}}\mathbf{I}\mathbf{U}_{\bm{\Gamma}}^T = \mathbf{U}_{\bm{\Gamma}}\mathbf{U}_{\bm{\Gamma}}^T$, which, clearly, is symmetric since
\begin{align}
\mathbf{\Theta}^T = (\mathbf{U}_{\bm{\Gamma}}\mathbf{U}_{\bm{\Gamma}}^T)^T = \mathbf{U}_{\bm{\Gamma}}\mathbf{U}_{\bm{\Gamma}}^T = \mathbf{\Theta}.
\end{align}
This completes the proof. \hfill $\square$

The projection through the Takagi factorization respects both the symmetry and unitary constraints required by the fully connected \ac{RIS}. For any invertible complex symmetric matrix with distinct singular values, the Takagi factorization can be implemented very efficiently by aligning the phases of the singular vectors computed via the \ac{SVD} \cite{Dieci22}.

\paragraph{Group-Connected Architecture}  
A similar strategy can be extended for a group-connected RIS architecture. In this case, the attacker must respect the group-wise constraints. Specifically, the reflection matrix $\mathbf{\Theta}$ should be block-diagonal, comprising $G$ independent sub-matrices $\mathbf{\Theta}_g$, for $g = 1, \cdots, G$, each one being symmetric and unitary. To generate such a matrix, the adversary follows the same process as for the fully connected RIS, but now for each group. First, it constructs random complex matrices $\tilde{\mathbf{\Gamma}}_g \in \mathbb{C}^{D_g \times D_g}$ for every group. Then, symmetrized versions $\mathbf{\Gamma}_g$ are achieved as in \eqref{eq:symmetrization}. Subsequently, the Takagi factorization is computed for each symmetrized matrix, as $ \mathbf{\Gamma}_g = \mathbf{U}_{\mathbf{\Gamma}_g} \mathbf{\Sigma}_{\mathbf{\Gamma}_g} \mathbf{U}_{\mathbf{\Gamma}_g}^T$. The symmetric unitary matrix for the $g$th group is then obtained via the projection $\mathbf{\Theta}_g = \mathbf{U}_{\mathbf{\Gamma}_g} \mathbf{U}_{\mathbf{\Gamma}_g}^T$, as in Lemma~\ref{lemma:uni_proj}. Finally, the full reflection matrix is assembled by organizing the $G$ sub-matrices into a block-diagonal structure, i.e., $\mathbf{\Theta} = \mathrm{bdiag}(\mathbf{\Theta}_1, \ldots, \mathbf{\Theta}_G)$.

\subsection{Aligned interference attack}\label{aligned_attack}

A more sophisticated adversarial strategy is to optimize the \ac{BD-RIS} reflection coefficients to maximize the interference induced at the users. As demonstrated in \cite{Sena24} for conventional RIS systems, if the adversary can acquire partial RIS-associated \ac{CSI}, it becomes possible to strategically steer the reflected power toward the users. In this subsection, we generalize this strategy to \ac{BD-RIS} architectures.

\paragraph{Fully Connected Architecture}
Let $\mu_i \in (0,1]$ denote the adversarial weight assigned to user $i \in \mathcal{U}$. The malicious fully connected \ac{RIS} controller aims to maximize the weighted sum of interference powers received at users, subject to unitary and symmetry constraints. Formally, the optimization problem can be formulated as:
\begin{subequations}\label{eq:opt_problem}
\begin{align}
\max_{\bm{\Theta}} \quad & \sum_{i \in \mathcal{U}} \mu_i \left\| \mathbf{g}_i^H \bm{\Theta} \mathbf{G} \right\|_2^2, \label{eq:opt_obj} \\
\text{s.t.} \quad & \bm{\Theta} = \bm{\Theta}^T. \label{eq:opt_problem_symmetry} \\
& \bm{\Theta}^H \bm{\Theta} = \mathbf{I}, \label{eq:opt_problem_unitary}
\end{align}
\end{subequations}

The unitary constraint in \eqref{eq:opt_problem_unitary} defines a non-convex Stiefel manifold. This, combined with the quadratic dependence of the objective function on $\bm{\Theta}$, makes the problem challenging to solve optimally. To simplify the original problem, we start by addressing the symmetry constraint in \eqref{eq:opt_problem_symmetry}. To this end, we exploit the mathematical structure of symmetric matrices by vectorizing $\bm{\Theta}$ using the concept of duplication matrix.

\property\label{prop:dup_mat}
For any symmetric matrix $\bm{\Theta}$, there exists a duplication matrix $\mathbf{D} \in \mathbb{R}^{D^2 \times \frac{D(D+1)}{2}}$ that maps the half-vectorization of $\bm{\Theta}$ to its full vectorization:
\begin{align}\label{dup_prop}
\mathrm{vec}(\bm{\Theta}) = \mathbf{D} \mathrm{vech}(\bm{\Theta}),    
\end{align}
where the duplication matrix $\mathbf{D}$ is uniquely defined as a sparse matrix with binary entries, given by:
\begin{align}
\mathbf{D}^T \triangleq \sum_{i=1}^{D} \sum_{j=1}^{i} \bm{u}_{ij} \mathrm{vec}(\mathbf{T}_{ij})^T,
\end{align}
where $\bm{u}_{ij}$ is a unit vector of length $\frac{D(D+1)}{2}$ with a single non-zero entry at position $[(j-1)D + i - \frac{j(j-1)}{2}]$, and $\mathbf{T}_{ij}$ is a $D \times D$ matrix with ones at positions $(i,j)$ and $(j,i)$, and zeros elsewhere \cite[Definitions 3.2a and 3.2b]{Magnus80}.

Property~\ref{prop:dup_mat} inherently satisfies the symmetry constraint of the original formulation, allowing us to reformulate the problem in a simpler relaxed version in terms of $\mathrm{vech}(\bm{\Theta})$.

\begin{figure}
\centering
\begingroup
\csname @twocolumnfalse\endcsname
\resizebox{.5\textwidth}{!}{%
\begin{minipage}{.6\textwidth}
\setlength{\algomargin}{5mm}
\setlength{\interspacetitleboxruled}{1mm}
{\begin{algorithm}[H]
\caption{Aligned Interference Attack for Fully Connected RIS}
\label{alg:adv_attack_fc}
\KwIn{Channel matrix $\mathbf{G}$, user channels $\{\mathbf{g}_i\}_{i\in\mathcal{U}}$, duplication matrix $\mathbf{D}$, user weights $\{\mu_i\}_{i\in\mathcal{U}}$}
\KwOut{Optimized reflection matrix $\bm{\Theta}^*$}

\vspace{1mm}

Compute the transformed interference channel matrices:

\ForEach{$i \in \mathcal{U}$}{
    Compute: $\mathbf{S}_i \leftarrow (\mathbf{G}^T \otimes \mathbf{g}_i^H)$\;
}
Construct the weighted interference matrix:
\[
\mathbf{S} \leftarrow [\sqrt{\mu_1}\mathbf{S}_1^H, \ldots, \sqrt{\mu_U}\mathbf{S}_{U}^H]^H.
\]

Compute the \ac{SVD} of $\bar{\bm{S}} = \mathbf{S}\mathbf{D}$ to solve the relaxed problem in \eqref{eq:stacked_problem}:
\[
\bar{\bm{S}} = \mathbf{U}_{\bar{\bm{S}}} \bm{\Sigma}_{\bar{\bm{S}}} \mathbf{V}_{\bar{\bm{S}}}^H.
\]

Obtain the optimized reduced coefficient vector:
$\bm{\theta} \leftarrow \mathbf{v}_{\bar{\bm{S}}, 1}$,
where $\mathbf{v}_{\bar{\bm{S}}, 1}$ is the corresponding dominant right singular vector.

Construct the relaxed reflection matrix:
$\bm{\Phi} \leftarrow \mathrm{unvec}(\mathbf{D} \bm{\theta})$.

Perform the Takagi factorization:
$\bm{\Phi} = \mathbf{U}_{\bm{\Phi}} \bm{\Sigma}_{\bm{\Phi}} \mathbf{U}_{\bm{\Phi}}^T$.

Project onto the set of unitary matrices:
$\bm{\Theta}^* \leftarrow \mathbf{U}_{\bm{\Phi}} \mathbf{U}_{\bm{\Phi}}^T$.

\Return $\bm{\Theta}^*$.
\end{algorithm}}
\end{minipage}
}%
\endgroup %
\end{figure}

Using the Kronecker product identity $\mathrm{vec}(\mathbf{A}\bm{\Theta}\mathbf{C}) = (\mathbf{C}^T \otimes \mathbf{A})\mathrm{vec}(\bm{\Theta})$, and invoking the duplication matrix relation in \eqref{dup_prop}, we can vectorize the interference channels $\mathbf{g}_i^H \bm{\Theta} \mathbf{G}$ as:
\begin{align}
\mathrm{vec}(\mathbf{g}_i^H \bm{\Theta} \mathbf{G})
&= (\mathbf{G}^T \otimes \mathbf{g}_i^H) \mathbf{D} \mathrm{vech}(\bm{\Theta}).
\end{align}
Then, by defining $\mathbf{S}_i = (\mathbf{G}^T \otimes \mathbf{g}_i^H) \in \mathbb{C}^{M \times D^2}$, and letting $\bm{\theta} = \mathrm{vech}(\bm{\Theta}) \in \mathbb{C}^{\frac{D(D+1)}{2}}$ denote the reduced vector of unique reflection coefficients, the objective function in \eqref{eq:opt_obj} becomes:
\begin{align}\label{sim_obj_func1}
\sum_{i \in \mathcal{U}} \mu_i &\left\| (\mathbf{G}^T \otimes \mathbf{g}_i^H) \mathbf{D} \bm{\theta} \right\|_2^2 \nonumber\\
& = \sum_{i \in \mathcal{U}} \mu_i \left\| \mathbf{S}_i \mathbf{D} \bm{\theta} \right\|_2^2\nonumber\\
& = \bm{\theta}^H \mathbf{D}^T \left(\sum_{i \in \mathcal{U}} \mu_i  \mathbf{S}_i^H \mathbf{S}_i \right)\mathbf{D} \bm{\theta}. 
\end{align}

\begin{figure}
\centering
\begingroup
\csname @twocolumnfalse\endcsname
\resizebox{.5\textwidth}{!}{%
\begin{minipage}{.6\textwidth}
\setlength{\algomargin}{5mm}
\setlength{\interspacetitleboxruled}{1mm}
{\begin{algorithm}[H]
\caption{Aligned Interference Attack for Group-Connected RIS}
\label{alg:adv_attack}
\KwIn{Channel matrix $\mathbf{G}$, user channels $\{\mathbf{g}_i\}_{i\in\mathcal{U}}$, \ac{RIS} group indices $\{\mathcal{E}_g\}_{g\in\mathcal{G}}$, duplication matrices $\{\mathbf{D}_g\}_{g\in\mathcal{G}}$, user weights $\{\mu_i\}_{i\in\mathcal{U}}$}
\KwOut{Optimized block-diagonal reflection matrix $\bm{\Theta}^*$}

Preprocess each RIS group:

\ForEach{$g \in \mathcal{G}$}{
  $D_g \leftarrow |\mathcal{E}_g|$\;
  Extract BS-RIS subchannel: $\mathbf{G}_g \leftarrow [\mathbf{G}]_{\mathcal{E}_g,:}$\;
  \ForEach{$i \in \mathcal{U}$}{
    Extract RIS-user subchannel: $\mathbf{g}_{ig} \leftarrow [\mathbf{g}_i]_{\mathcal{E}_g}$\;
    Compute: $\mathbf{S}_{ig} \leftarrow \bigl(\mathbf{G}_g^T \otimes \mathbf{g}_{ig}^H\bigr)$\;
  }
}

Construct the interference matrix from group contributions:

\ForEach{$i \in \mathcal{U}$}{
  Compute the block-row matrix:
  \[
  \mathbf{J}_i \leftarrow \begin{bmatrix} \mathbf{S}_{i1}\mathbf{D}_1 & \mathbf{S}_{i2}\mathbf{D}_2 & \ldots & \mathbf{S}_{iG}\mathbf{D}_G \end{bmatrix}\,.
  \]
}

Construct the stacked global matrix:
\[
\mathbf{J} \leftarrow 
\begin{bmatrix} 
\sqrt{\mu_1} \mathbf{J}_1^T, 
\sqrt{\mu_2} \mathbf{J}_2^T, 
\ldots, 
\sqrt{\mu_{U}} \mathbf{J}_{U}^T 
\end{bmatrix}^T\,.
\]

Compute the \ac{SVD} of $\mathbf{J}$ to solve the relaxed problem in \eqref{prob_rex_gRIS2}:
\[
\mathbf{J} = \mathbf{U}_\mathbf{J}\bm{\Sigma}_\mathbf{J} \mathbf{V}_\mathbf{J}^H\,.
\]

Obtain the optimized coefficient vector:
$\bm{\varphi} \leftarrow \mathbf{v}_{\mathbf{J}, 1}$, 
where $\mathbf{v}_{\mathbf{J}, 1}$ is the corresponding dominant right singular vector.

\ForEach{$g \in \mathcal{G}$}{
  Extract reduced vector:
  $\bm{\theta}_g \leftarrow [\bm{\varphi}]_{\mathcal{R}_g}$,
  where $\mathcal{R}_g \subset \left\{1, \ldots, \sum_{g\in\mathcal{G}}\frac{D_g(D_g+1)}{2}\right\}$ is the index set for group $g$, with $|\mathcal{R}_g|=\frac{D_g(D_g+1)}{2}$.
  
 Construct the relaxed reflection matrix:
  $\bm{\Phi}_g \leftarrow \mathrm{unvec}(\mathbf{D}_g\,\bm{\theta}_g)$.
  
  Perform the Takagi factorization:
  $\bm{\Phi}_g =  \mathbf{U}_{\bm{\Phi}_g} \bm{\Sigma}_{\bm{\Phi}_g} \mathbf{U}_{\bm{\Phi}_g}^T$.
  
  Project onto the set of unitary matrices:
  $\bm{\Theta}^*_g \leftarrow \mathbf{U}_{\bm{\Phi}_g} \mathbf{U}_{\bm{\Phi}_g}^T.$
}

Construct the final block-diagonal reflection matrix:
\[
\bm{\Theta}^* \leftarrow \mathrm{bdiag}\bigl(\bm{\Theta}^*_1,\,\bm{\Theta}^*_2,\,\dots,\,\bm{\Theta}^*_G\bigr).
\]

\Return $\bm{\Theta}^*$.
\end{algorithm}}
\end{minipage}
}%
\endgroup %
\end{figure}

Given that $\sum_{i \in \mathcal{U}}  \mathbf{S}_i^H \mathbf{S}_i = \left[\mathbf{S}_1^H, \ldots, \mathbf{S}_{U}^H\right]^H \left[\mathbf{S}_1^H, \cdots, \mathbf{S}_{U}^H\right]$, we can define the tall weighted interference matrix: $\mathbf{S} \triangleq [\sqrt{\mu_1}\mathbf{S}_1^H, \ldots, \sqrt{\mu_U}\mathbf{S}_{U}^H]^H \in \mathbb{C}^{U N \times D^2}$ and rewrite the expression in \eqref{sim_obj_func1} as the convex quadratic objective $\bm{\theta}^H \mathbf{D}^T \mathbf{S}^H \mathbf{S} \mathbf{D} \bm{\theta}$. This vectorized reformulation automatically satisfies the symmetry constraint in \eqref{eq:opt_problem_symmetry} by construction. On the other hand, the non-convex unitary constraint in \eqref{eq:opt_problem_unitary} is relaxed with a convex $\ell_2$-norm bound $\|\bm{\theta}\|_2\leq1$, which allows us to formulate a tractable optimization problem:
\begin{subequations}\label{eq:stacked_problem}  
\begin{align}  
\max_{\bm{\theta}} \quad & \bm{\theta}^H \mathbf{D}^T \mathbf{S}^H \mathbf{S} \mathbf{D} \bm{\theta} \label{eq:simp_prob_obj} \\  
\text{s.t.} \quad & \|\bm{\theta}\|_2 \leq 1. \label{eq:stacked_problem_b}  
\end{align}
\end{subequations}

This relaxed problem \eqref{eq:stacked_problem} constitutes a \ac{QCQP}, whose optimal solution is the dominant right singular vector of the matrix $\bar{\bm{S}} \triangleq \mathbf{S} \mathbf{D}$. More specifically, by recalling the \ac{SVD}, we can express $\bar{\bm{S}} = \mathbf{U}_{\bar{\bm{S}}} \bm{\Sigma}_{\bar{\bm{S}}} \mathbf{V}_{\bar{\bm{S}}}^H$, where $\mathbf{V}_{\bar{\bm{S}}} = [\mathbf{v}_{\bar{\bm{S}}, 1}, \mathbf{v}_{\bar{\bm{S}}, 2}, \dots, \mathbf{v}_{\bar{\bm{S}},\frac{D(D+1)}{2}}]$ contains the right singular vectors. Then, the optimal solution to the relaxed problem is given by $\bm{\theta} = \mathbf{v}_{\bar{\bm{S}}, 1}$, thus satisfying $\|\bm{\theta}\|_2 = 1$, and the corresponding relaxed \ac{BD-RIS} matrix is obtained by:
\begin{align}
\bm{\Phi} = \mathrm{unvec}(\mathbf{D} \bm{\theta}).
\end{align}
However, $\bm{\Phi}$ generally violates the unitary constraint, i.e., $\bm{\Phi}^H \bm{\Phi} \neq \mathbf{I}$. To enforce this, we project $\bm{\Phi}$ onto the closest unitary matrix via the Takagi factorization explained in Property~\ref{prop:takagi}, i.e., $\bm{\Phi} = \mathbf{U}_{\bm{\Phi}} \bm{\Sigma}_{\bm{\Phi}} \mathbf{U}_{\bm{\Phi}}^T$, allowing us to compute
\begin{align}\label{projection}
 \bm{\Theta}^* = \mathbf{U}_{\bm{\Phi}} \mathbf{U}_{\bm{\Phi}}^T,
\end{align}
where, according to Lemma~\ref{lemma:uni_proj}, the symmetry is preserved.

Note that relaxing the unitary constraint in \eqref{eq:stacked_problem} provides a tractable surrogate formulation of \eqref{eq:opt_problem}, whose solution captures the dominant interference direction of the original problem via \ac{SVD}. Subsequently, the projection in \eqref{projection} returns the closest symmetric unitary matrix in the Frobenius norm sense, as per Lemma~\ref{lemma:uni_proj}, to the relaxed solution, and thus enforces feasibility but does not guarantee global optimality. Since \eqref{eq:opt_problem} is inherently non-convex and its global optimum is generally intractable to characterize, determining the exact performance gap between the proposed suboptimal adversarial approach and the original formulation remains an open challenge.

\paragraph{Group-Connected Architecture}
\label{subsec:group_opt}

Herein, the aligned interference attack is specialized to the group-connected architecture. Let us consider a \ac{BD-RIS} partitioned into $G$ independent groups, i.e., $\bm{\Theta} = \mathrm{bdiag}(\bm{\Theta}_1, \dots, \bm{\Theta}_G)$, with each block satisfying $\bm{\Theta}_g = \bm{\Theta}_g^T$ and $\bm{\Theta}_g^H\bm{\Theta}_g = \mathbf{I}$. Then, the optimized adversary attack strategy can be formulated as:
\begin{subequations}\label{eq:opt_problem2}
\begin{align}
\max_{\bm{\Theta}} \quad & \sum_{i \in \mathcal{U}} \mu_i \left\| \mathbf{g}_i^H \bm{\Theta} \mathbf{G} \right\|_2^2, \label{eq:opt2_obj} \\
\text{s.t.} \quad & \bm{\Theta} = \mathrm{bdiag}(\bm{\Theta}_1, \dots, \bm{\Theta}_G),\\
& \bm{\Theta}_g = \bm{\Theta}_g^T, \label{eq:opt_problem2_symmetry} \\
& \bm{\Theta}_g^H \bm{\Theta}_g = \mathbf{I}. \label{eq:opt_problem2_unitary}
\end{align}
\end{subequations}

To address problem~\ref{eq:opt_problem2}, the first step is to decompose the global channel matrices into group-specific subchannels. To this end, let $\mathcal{E}_g \subset \{1,\dots, D\}$ denote the index subset of RIS elements associated with group $g$, with $|\mathcal{E}_g|=D_g$. Then, the BS-RIS subchannel for group $g$ can be expressed as
\begin{align}
    \mathbf{G}_g = [\mathbf{G}]_{\mathcal{E}_g,:} \in \mathbb{C}^{D_g \times M},
\end{align}
while the RIS-user subchannel becomes
\begin{align}
    \mathbf{g}_{ig} = [\mathbf{g}_i]_{\mathcal{E}_g} \in \mathbb{C}^{D_g}.
\end{align}

Next, following Property~\ref{prop:dup_mat}, we construct a duplication matrix $\mathbf{D}_g \in \mathbb{R}^{D_g^2 \times \frac{D_g(D_g+1)}{2}}$ to vectorize the symmetric matrix as $\mathrm{vec}(\bm{\Theta}_g) = \mathbf{D}_g\bm{\theta}_g$ for each group, where $\bm{\theta}_g = \mathrm{vech}(\bm{\Theta}_g) \in \mathbb{C}^{\frac{D_g(D_g+1)}{2}}$ denotes the $g$th reduced coefficient vector. We also define $\mathbf{S}_{ig} = (\mathbf{G}_g^T \otimes \mathbf{g}_{ig}^H) \in \mathbb{C}^{M \times D_g^2}$, and represent the set of RIS groups by $\mathcal{G} = \{1, \ldots, G\}$. Then, we rewrite the adversary objective as follows:
\begin{align}\label{sim_obj2_func1}
\sum_{i \in \mathcal{U}} \mu_i &\left\| \sum_{j\in\mathcal{G}}(\mathbf{G}_j^T \otimes \mathbf{g}_{ij}^H) \mathbf{D}_j \bm{\theta}_j \right\|_2^2 \nonumber\\
&\hspace{10mm}= \sum_{i \in \mathcal{U}} \mu_i \left\| \sum_{j\in\mathcal{G}}\mathbf{S}_{ij} \mathbf{D}_j \bm{\theta}_j \right\|_2^2. 
\end{align}

To further simplify, we introduce a global stacked vector:
\begin{align}
\bm{\varphi} \triangleq 
[\bm{\theta}_{1}^T, \bm{\theta}_{2}^T, \ldots, \bm{\theta}_{G}^T]^T \in \mathbb{C}^{\sum_{g\in \mathcal{G}}\frac{D_g(D_g+1)}{2}},
\end{align}
and define, for each user $i \in \mathcal{U}$, the block row matrix:
\begin{align}
\mathbf{J}_i \triangleq 
\begin{bmatrix} 
\mathbf{S}_{i1}\mathbf{D}_1, \mathbf{S}_{i2}\mathbf{D}_2, \ldots, 
\mathbf{S}_{iG}\mathbf{D}_G
\end{bmatrix}.
\end{align}

Now, the summation related to the groups can be written as a single matrix structure, as follows:
\begin{align}\label{prefinal_g_obj}
 \sum_{i \in \mathcal{U}} \mu_i &\left\| \sum_{j\in\mathcal{G}}\mathbf{S}_{ij} \mathbf{D}_j \bm{\theta}_j \right\|_2^2 \nonumber\\
 &= \bm{\varphi}^H \left( \sum_{i \in \mathcal{U}} \mu_i \mathbf{J}_i^H \mathbf{J}_i \right) \bm{\varphi}.
\end{align}

Then, we define a global matrix $\mathbf{J}$ stacking all $\mathbf{J}_i$ matrices:
\begin{align}
\mathbf{J} \triangleq 
\begin{bmatrix} 
\sqrt{\mu_1} \mathbf{J}_1^T, 
\sqrt{\mu_2} \mathbf{J}_2^T, 
\ldots, 
\sqrt{\mu_{U}} \mathbf{J}_{U}^T 
\end{bmatrix}^T.
\end{align}

This allows us to write the summation in \eqref{prefinal_g_obj} in its matrix-equivalent form, resulting in the final compact problem:
\begin{subequations}\label{prob_rex_gRIS2}
\begin{align}
    \underset{{\bm{\varphi}}}{\max} \hspace{3mm} & 
         \bm{\varphi}^H\mathbf{J}^H \mathbf{J} \bm{\varphi},\\[0mm]
    \text{s.t.} \hspace{4mm}& \|{\bm{\varphi}}\|_2 \leq 1.
\end{align}
\end{subequations}

\begin{table*}[t]
\centering
\caption{Runtime (seconds) of optimized attacks for different RIS architectures}
\label{tab:runtime}
\renewcommand{\arraystretch}{1.2}
\begin{tabular}{l|cccccc|}
\cline{2-7}
\multicolumn{1}{c}{} & \multicolumn{6}{|c|}{Number of RIS elements $D$} \\
\cline{2-7}
\multicolumn{1}{c}{} & \multicolumn{1}{|c}{50} & 100 & 150 & 200 & 250 & 300 \\
\hline
\multicolumn{1}{|l|}{Fully Connected RIS (proposed, Algorithm~\ref{alg:adv_attack_fc})} & 0.20198 & 0.49007 & 0.89702 & 1.14418 & 1.84436 & 2.36110 \\
\multicolumn{1}{|l|}{Group-Connected RIS (proposed, Algorithm~\ref{alg:adv_attack})} & 0.04219 & 0.05327 & 0.09056 & 0.10905 & 0.11455 & 0.13659 \\
\multicolumn{1}{|l|}{Single-Connected RIS (iterative baseline~\cite{Sena24})} & 0.70900 & 1.11176 & 1.36038 & 1.52885 & 1.82989 & 2.08470 \\
\hline
\end{tabular}
\end{table*}

The solution to \eqref{prob_rex_gRIS2} can be obtained with the aid of the SVD, similarly to the problem \eqref{eq:stacked_problem}. More specifically, we can decompose $\mathbf{J} = \mathbf{U}_\mathbf{J}\bm{\Sigma}_\mathbf{J} \mathbf{V}_\mathbf{J}^H$ and achieve the desired solution by computing $\bm{\varphi} = \mathbf{v}_{\mathbf{J},1}$, where $\mathbf{v}_{\mathbf{J},1}$ is the dominant right singular vector of $\mathbf{J}$, i.e., the first column of $\mathbf{V}_{\mathbf{J}}$.
Finally, by letting $\mathcal{R}_g \subset \left\{1, \ldots, \sum_{g \in \mathcal{G}}\frac{D_g(D_g+1)}{2} \right\}$ denote the index subset corresponding to the $g$th reduced coefficient vector, with $|\mathcal{R}_g| = \frac{D_g(D_g+1)}{2}$, the relaxed coefficients for each independent \ac{RIS} group can be obtained as $\bm{\theta}_{g} = [\bm{\varphi}]_{\mathcal{R}_g}$, and $\bm{\Phi}_{g} = \mathrm{unvec}(\mathbf{D}_g\bm{\theta}_{g})$, $\forall g \in \mathcal{G}$. Recalling Property~\ref{prop:takagi}, we compute the Takagi factorization $\bm{\Phi}_g = \mathbf{U}_{\bm{\Phi}_g} \bm{\Sigma}_{\bm{\Phi}_g} \mathbf{U}_{\bm{\Phi}_g}^T$ and project each $\bm{\Phi}_g$ onto its closest unitary matrix, as follows:
\begin{align}
 \bm{\Theta}^*_g = \mathbf{U}_{\bm{\Phi}_g} \mathbf{U}_{\bm{\Phi}_g}^T.
\end{align} 

Finally, the full BD-RIS reflection matrix is constructed as $\bm{\Theta}^* = \mathrm{bdiag}(\bm{\Theta}^*_1, \ldots, \bm{\Theta}^*_G)$.

\subsection{Complexity Analysis}\label{subsec:complexity}

In this subsection, we analyze the computational complexity of the proposed \ac{BD-RIS}-induced attacks, in terms of complex \acp{FLOP}, quantifying their scaling with key system parameters, such as the number of RIS elements $D$, transmit antennas $M$, and users $U$.

\paragraph{Random Interference Attack} 
The complexity load of this strategy is dominated by the generation of valid reflection matrices, as explained in Section~\ref{subsec:random_interf}.
For the fully connected \ac{RIS}, the procedure begins by generating and symmetrizing a $D \times D$ complex matrix $\mathbf{\Gamma}$, which requires $\mathcal{O}(D^2)$ complex \acp{FLOP}.
The dominant cost is the projection step, performed via the Takagi factorization. This is implemented through the \ac{SVD} of $\mathbf{\Gamma}$. Specifically, the standard \ac{SVD} of an $A \times B$ matrix has complexity $\mathcal{O}(AB\min(A,B))$~\cite{Halko2011}. Since $\mathbf{\Gamma}$ is $D\times D$, the associated cost is $\mathcal{O}(D^3)$.
The subsequent matrix multiplication, $\mathbf{\Theta} = \mathbf{U}_{\bm{\Gamma}}\mathbf{U}_{\bm{\Gamma}}^T$, also requires $\mathcal{O}(D^3)$. Therefore, the total complexity is $\mathcal{O}(D^2 + D^3 + D^3)$, yielding a worst-case complexity of $\mathcal{O}(D^3)$.

For the group-connected \ac{RIS}, the same procedure is carried out independently for each group. For the $g$th group, the generation and symmetrization of $\mathbf{\Gamma}_g$ requires $\mathcal{O}(D_g^2)$, while the subsequent \ac{SVD} and the final matrix multiplication require $\mathcal{O}(D_g^3)$. The total complexity is therefore $\sum_{g=1}^{G}\mathcal{O}(D_g^2 + D_g^3 + D_g^3)$. In the case where groups are equal-sized, i.e., $D_g = D/G$, the worst-case complexity becomes $\mathcal{O}(D^3/G^2)$.

\paragraph{Aligned Interference Attack}
Algorithm~\ref{alg:adv_attack_fc}, for the fully connected \ac{RIS}, begins by constructing the interference matrices (lines 1--4), which involves the computation of $U$ Kronecker products, yielding $\mathcal{O}(UMD^2)$ complex \acp{FLOP}. The subsequent multiplication $\bar{\mathbf{S}} = \mathbf{S}\mathbf{D}$ (line 5) can also be executed in $\mathcal{O}(UMD^2)$ \acp{FLOP} when exploiting the sparse structure of $\mathbf{D}$.
The most expensive operation is the \ac{SVD} of $\bar{\mathbf{S}} \in \mathbb{C}^{UM \times {D(D+1)}/{2}}$. When the \ac{RIS} is relatively small, such that $D(D+1)/2 < UM$, the matrix $\bar{\mathbf{S}}$ is tall and its \ac{SVD} can be computed in $\mathcal{O}\!\left(UM\left({D(D+1)}/{2}\right)^2\right) \approx \mathcal{O}(UMD^4)$ \acp{FLOP}. The total complexity is thus $\mathcal{O}(UMD^2 + UMD^4 + D^3)$, with a worst-case complexity of $\mathcal{O}(UMD^4)$. 
On the other hand, for large \acp{RIS} where $D(D+1)/2 > UM$, $\bar{\mathbf{S}}$ is wide, leading to a complexity of $\mathcal{O}\!\left(U^2M^2{D(D+1)}/{2}\right) \approx \mathcal{O}(U^2M^2D^2)$. The total complexity in this regime is $\mathcal{O}(UMD^2 + U^2M^2D^2 + D^3)$, with a worst-case complexity of $\mathcal{O}(U^2M^2D^2)$.

Algorithm~\ref{alg:adv_attack}, for the group-connected \ac{RIS}, starts with matrix transformations and Kronecker products (lines 1--11), with a combined cost of $\mathcal{O}\!\left(UM\sum_{g=1}^{G}D_g^2\right)$ complex \acp{FLOP}. The complexity for the \ac{SVD} of $\mathbf{J} \in \mathbb{C}^{UM \times \sum_{g=1}^{G} \frac{D_g(D_g+1)}{2}}$ (line 12) dominates this strategy. When $\mathbf{J}$ is tall, i.e., its number of rows is greater than its number of columns, the \ac{SVD} requires $\mathcal{O}\!\left(UM\left(\sum_{g=1}^{G}{D_g(D_g+1)}/{2}\right)^2\right)$ \acp{FLOP}, whereas for a wide matrix $\mathbf{J}$ the complexity becomes $\mathcal{O}\!\left(U^2M^2\sum_{g=1}^{G}{D_g(D_g+1)}/{2}\right)$. The final projections (lines 14--18) require $G$ Takagi factorizations, resulting in $\sum_{g=1}^{G}\mathcal{O}(D_g^3)$ \acp{FLOP}.
For equally-sized groups, the worst-case complexities for the tall and wide cases of $\mathbf{J}$ reduce to $\mathcal{O}(UMD^4/G^2)$ and $\mathcal{O}(U^2M^2D^2/G)$, respectively.

The analysis for the random strategy shows that \acp{BD-RIS} impose a greater computational load compared to a conventional single-connected \ac{RIS}, which has a linear complexity of $\mathcal{O}(D)$. For the aligned attack with single-connected \acp{RIS}, iterative optimization methods are often employed~\cite{Sena24, rivetti2024malicious}, which typically have lower per-iteration complexity but require multiple iterations for convergence. In contrast, Algorithms~\ref{alg:adv_attack_fc} and~\ref{alg:adv_attack} are non-iterative, and their computational cost is primarily driven by the dimensionality of the considered BD-RIS architectures.
To illustrate this behavior, Table~\ref{tab:runtime} reports the average runtime of the considered attack strategies for different numbers of RIS elements. The results show that the runtime of all methods increases with the RIS size, in agreement with the complexity scaling trends discussed earlier. The single-connected RIS baseline based on the iterative optimization in \cite{Sena24} presents higher runtime values up to $D=200$. However, its lower per-iteration complexity enables it to become more competitive for larger RIS sizes (i.e., $D\geq250$), despite its iterative nature. In this large-scale regime, the fully connected RIS becomes the most computationally demanding configuration, consistent with its quartic scaling in $D$.
On the other hand, the group-connected RIS (with a group size fixed to five elements) exhibits the lowest runtime values across all evaluated configurations, reflecting the reduced problem dimension achieved by its partitioned architecture.

The above analysis provides insight into the practical implications of the proposed attacks. On the one hand, the worst-case complexity of $\mathcal{O}(UMD^4)$ for the fully connected RIS, which is reflected in the runtimes reported in Table~\ref{tab:runtime}, indicates that scaling fully connected architectures to very large sizes may become challenging. On the other hand, the group-connected RIS significantly reduces the computational burden, thus offering a more scalable BD-RIS implementation, while still causing a strong adversarial impact, as will be seen in Section~\ref{res_sec}.

\begin{table}[t]
\centering
\caption{Summary of simulation parameters}
\label{tab:sim_params}
\renewcommand{\arraystretch}{1.2}
\begin{tabular}{|l|l|}
    \hline
    Parameter & Value \\
    \hline
    Number of BS antennas $M$ & $32$ \\
    Number of RIS elements $D$ & $200$ \\
    Group-connected RIS size $D_g$ & $5$ \\
    Number of users $U$ & $3$ \\
    User distances $\{d_u\}$ & $\{30, 50, 60\}$ m \\
    User azimuth angles $\{\theta_u\}$ & $\{25^\circ, 15^\circ, 10^\circ\}$ \\
    RIS distance $d_{\text{RIS}}$ & $40$ m \\
    RIS azimuth angle $\theta_{\text{RIS}}$ & $5^\circ$ \\
    Path-loss exponent $\eta$ & $3$ \\
    CSI error factors & $\epsilon^{\text{\tiny BS-U}}, \epsilon^{\text{\tiny BS-RIS}},  \epsilon^{\text{\tiny RIS-U}} \in [0, 0.7]$ \\
    SIC error factor & $\xi \in [0, 7\times 10^{-3}]$ \\
    Noise power & $-60$ dBm \\
    \hline
    \end{tabular}
\end{table}

\section{Numerical Results}\label{res_sec}

In this section, we evaluate the impact of \ac{BD-RIS}-induced attacks under both random and optimized strategies and investigate the ability of \ac{RSMA} to maintain performance under adversarial conditions. We analyze the degradations induced by fully connected, group-connected, and conventional single-connected \ac{RIS} architectures and provide a comprehensive performance comparison with \ac{SDMA}.

The simulation setup comprises: (i) a \ac{BS} equipped with $M = 32$ transmit antennas positioned at the origin; (ii) three users located at distances of $d_1 = 30$~m, $d_2 = 50$~m, and $d_3 = 60$~m from the \ac{BS}, with azimuth angles of $\theta_1 = 25^\circ$, $\theta_2 = 15^\circ$, and $\theta_3 = 10^\circ$; and (iii) an adversarial \ac{RIS} deployed at a distance of $d_{\text{RIS}} = 40$~m from the \ac{BS} with an azimuth angle of $\theta_{\text{RIS}} = 5^\circ$.
Unless stated otherwise, simulations are performed for $D = 200$ reflecting elements, with the group-connected \ac{RIS} configured as $D_1 = \cdots = D_G = 5$, considering a standard imperfect \ac{CSI} scenario with $ \epsilon^{\text{\tiny BS-U}} = \epsilon^{\text{\tiny BS-RIS}} = \epsilon^{\text{\tiny RIS-U}} = 0.3$.
The small-scale fading for all channel links is modeled following a Rayleigh distribution, representing a rich scattering environment, combined with a power-law path loss model $d^{-\eta}$ with $\eta = 3$, where $d$ denotes the distance. In particular, the distance from the \ac{RIS} to the $u$-th user is obtained by $d_{\text{RIS},u} = \sqrt{d_{\text{RIS}}^2 + d_u^2 - 2 d_{\text{RIS}}d_u \cos(\theta_{\text{RIS}} - \theta_u)}$, resulting in $d_{\text{RIS},1} \approx 15.64$~m, $d_{\text{RIS},2} \approx 12.68$~m, and $d_{\text{RIS},3} \approx 20.45$~m. Moreover, the user weights for Algorithms~\ref{alg:adv_attack_fc} and~\ref{alg:adv_attack} are set uniformly as $\mu_1 = \mu_2 =\mu_3 = 1/3$, the noise power is set to $-60$~dBm, \ac{SDMA} employs a uniform power allocation with regularized zero-forcing precoders, and \ac{RSMA} operates under perfect SIC by default. Table~\ref{tab:sim_params} summarizes simulated parameter values. 
The sensitivity to \ac{SIC} imperfections, as well as the impact of varying \ac{CSI} errors, is investigated in Section~\ref{subsec:imperfections}.

\subsection{Impact of the RIS Uplink Mode}\label{subsec:uplink_modes}

\begin{figure}[t]
	\centering
	\includegraphics[width=1.0\linewidth]{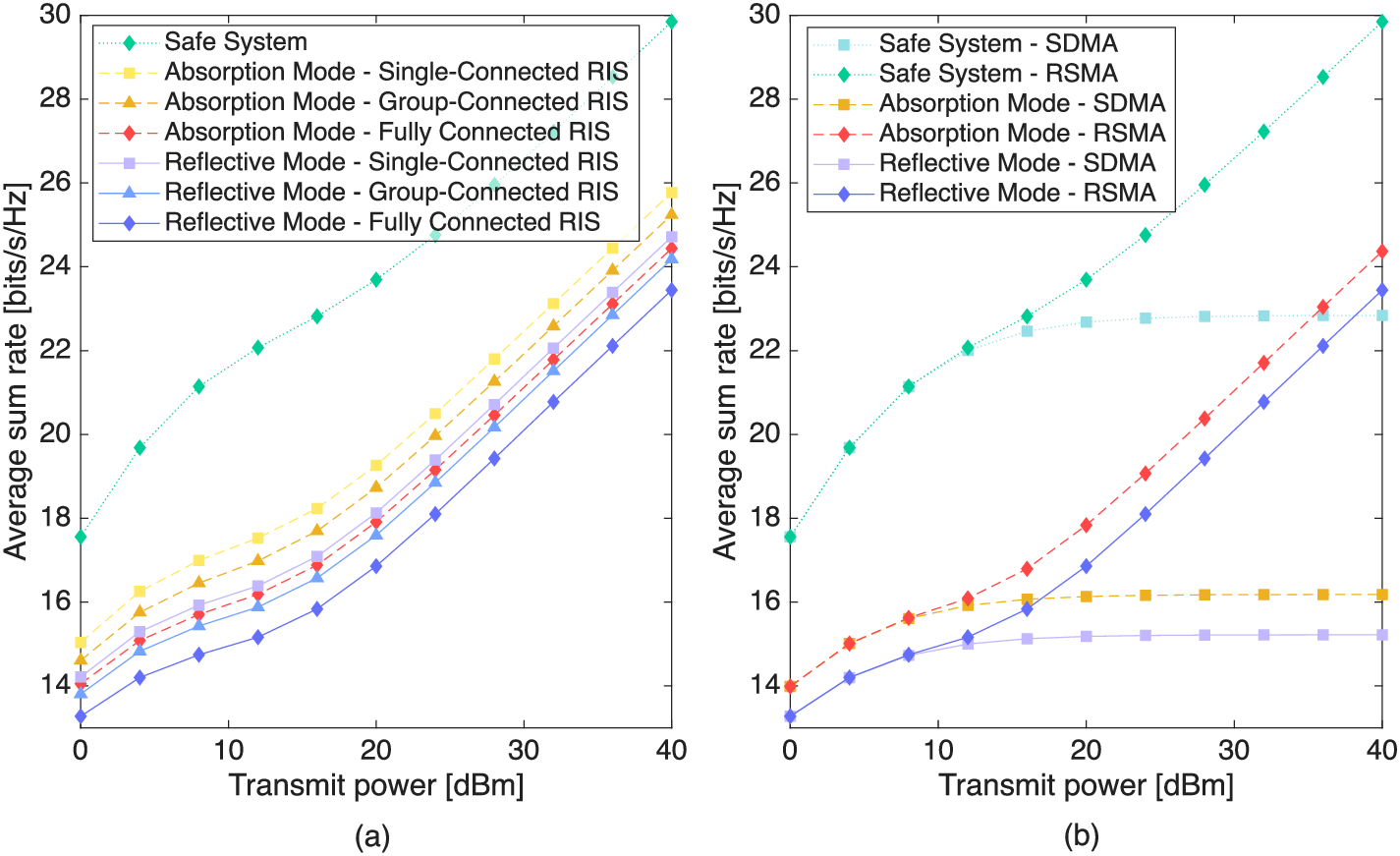}
	\caption{Impact of the \ac{RIS} mode in the uplink training phase: (a) average sum rate of \ac{RSMA} across different \ac{RIS} architectures; and (b) comparison between \ac{RSMA} and \ac{SDMA} under the fully connected \ac{RIS}. Results are shown for the aligned interference attack.}\label{fig:res_1}
\end{figure}

Fig.~\ref{fig:res_1} investigates the impact of the \ac{RIS} operation mode during the uplink training phase on the system average sum rate under aligned interference attacks. Fig.~\ref{fig:res_1}(a) evaluates this impact across different \ac{RIS} architectures for the \ac{RSMA} scheme. It can be seen that using a reflective mode during uplink training significantly boosts the degradation potential across all architectures, but the benefit to the attacker is greatest when using \acp{BD-RIS}. The fully connected \ac{RIS}, for instance, reduces the sum rate to $23.4$~bits/s/Hz at $40$~dBm, a more severe degradation compared to the $24.7$~bits/s/Hz achieved by the single-connected \ac{RIS}.

Fig.~\ref{fig:res_1}(b) then provides a direct comparison between \ac{RSMA} and \ac{SDMA} for the most potent fully connected architecture. The results confirm that the reflective mode is more damaging than the absorption mode for both multiple access schemes. The figure also sheds light on \ac{RSMA}'s inherent robustness, as it consistently outperforms \ac{SDMA} across all attack scenarios. This superior performance is a result of the adaptive power allocation described in Section~\ref{pa_subsection}, which enables \ac{RSMA} to better manage channel estimation errors induced by the reflective mode attack.
These observations provide the first evidence of \ac{RSMA}'s robustness advantage over \ac{SDMA}, which will be further analyzed in the next subsection using the proposed robustness metrics.

\begin{figure}[t]
	\centering
	\includegraphics[width=1.0\linewidth]{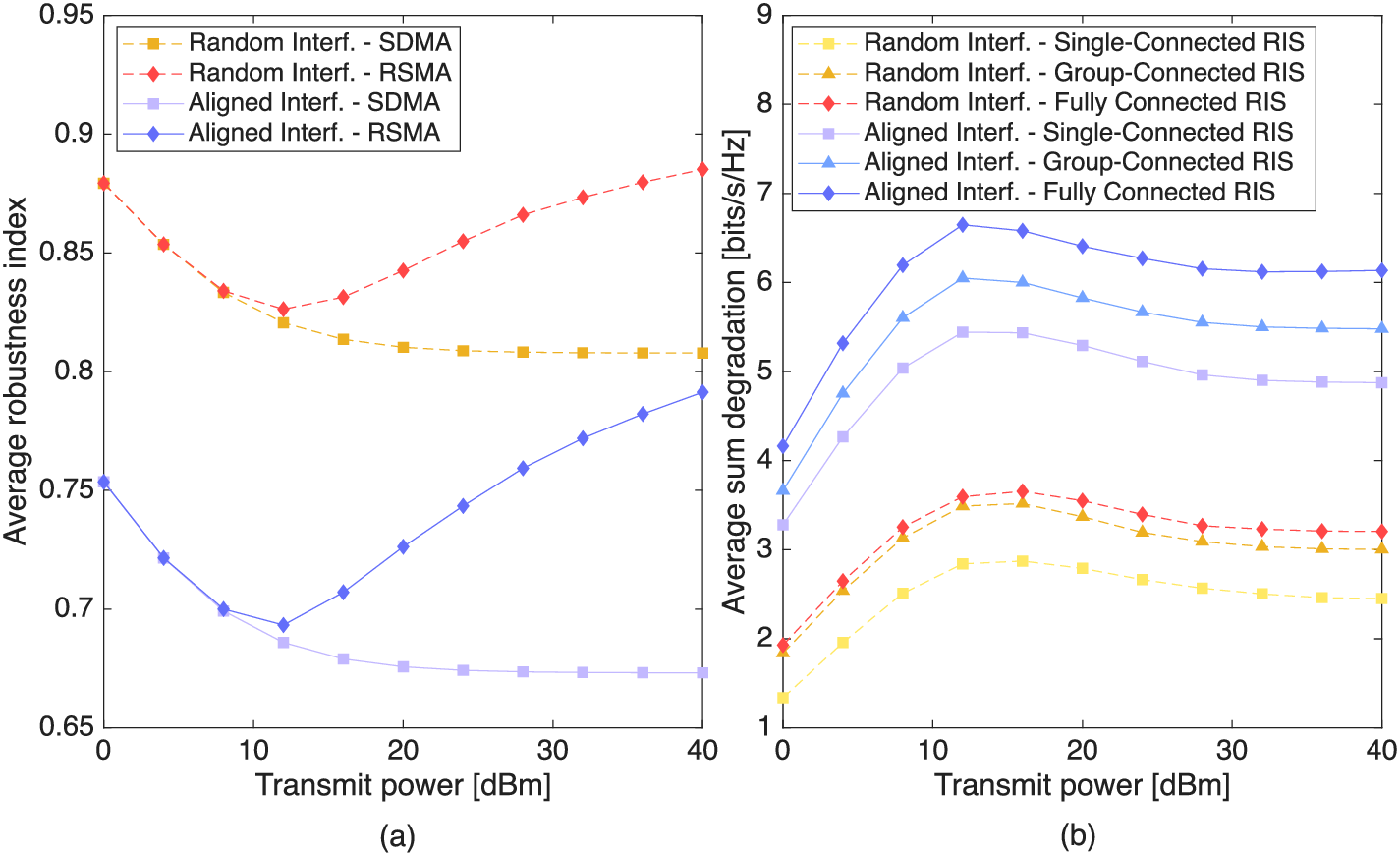}
	\caption{Performance degradation and robustness analysis: (a) average robustness index comparing \ac{RSMA} and \ac{SDMA} under the fully connected \ac{RIS}; and (b) average sum-rate degradation of \ac{RSMA} under random and aligned attacks across different \ac{RIS} architectures.}\label{fig:res_3}
\end{figure}

\begin{figure*}[t]
	\centering
	\includegraphics[width=.76\linewidth]{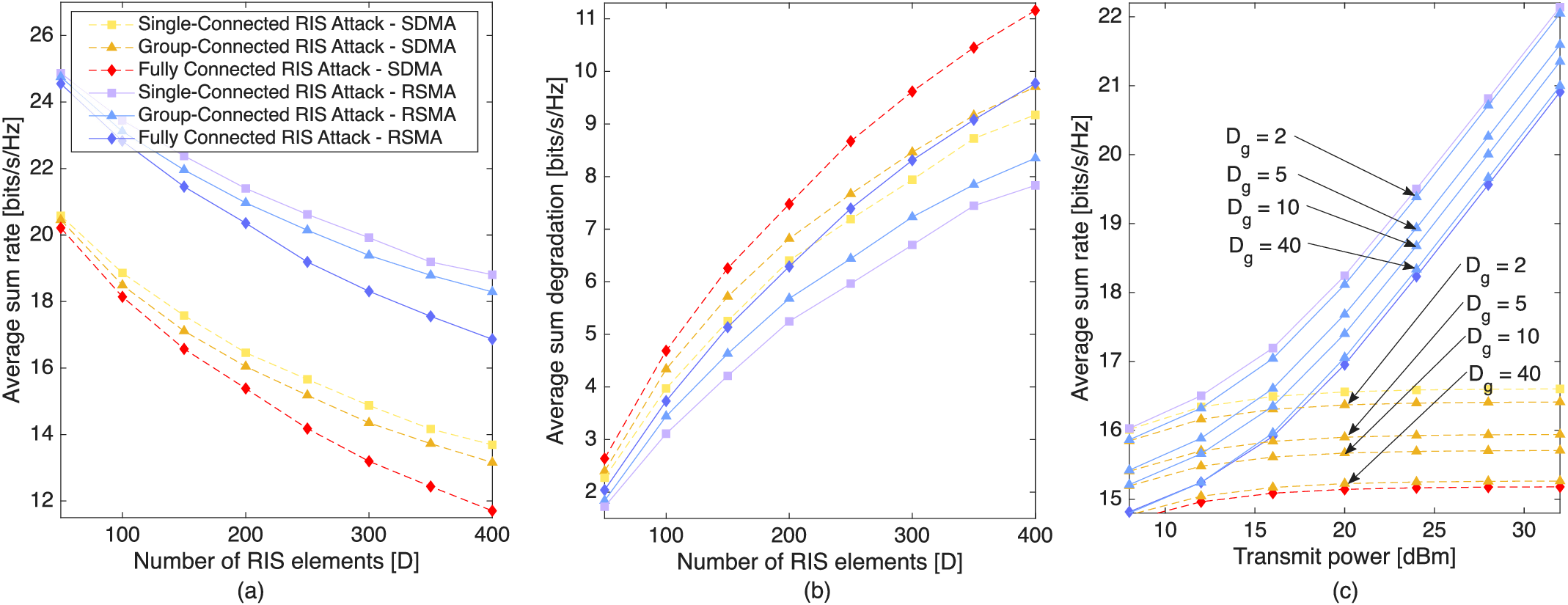}
	\caption{Impact of \ac{RIS} size and architecture parameters: (a) average sum rate versus $D$; (b) average sum-rate degradation versus $D$; and (c) average sum rate versus transmit power for different group sizes $D_g$ of the group-connected \ac{RIS}. In (a) and (b), the group-connected \ac{RIS} uses the default group size $D_g = 5$ and a fixed transmit power of $30$~dBm. Results assume the aligned interference attack.}\label{fig:res_4}
\end{figure*}

\subsection{Robustness of RSMA}

Given that operating in reflective mode during uplink pilot training leads to more severe degradation, we adopt this mode in the subsequent results. The analysis now turns to the robustness of \ac{RSMA} in comparison to \ac{SDMA} and across different \ac{RIS} architectures.
Fig.~\ref{fig:res_3}(a) presents the robustness index under the fully connected \ac{RIS}. The results show that \ac{RSMA} consistently maintains a significantly higher robustness index than \ac{SDMA} under both random and aligned attacks.
For example, under the random interference attack at $40$~dBm, \ac{RSMA} achieves an index of about $0.88$, whereas \ac{SDMA} drops to $0.81$. Under the aligned attack, \ac{RSMA} still recovers to $0.79$ at $40$~dBm, while \ac{SDMA} reaches only $0.67$.
This observation indicates that the gap in robustness between the two schemes widens under the aligned attack, highlighting that \ac{RSMA}'s performance advantage is even larger when facing severe interference. It is also clear that the aligned attack is consistently more harmful than the random interference, validating the proposed attack strategies.

Another important characteristic of the robustness curves in Fig.~\ref{fig:res_3}(a) is the non-monotonic trend of \ac{RSMA}, with a noticeable rebound after $12$~dBm. The origin of this behavior becomes clear when jointly examining the robustness curves and the corresponding sum-rate degradations of \ac{RSMA} across different \ac{RIS} architectures in Fig.~\ref{fig:res_3}(b). The degradation curves directly quantify the relative gap between the sum rates of the attacked and safe systems, i.e., $\sum_{\forall u} \Delta R_u$, as defined in \eqref{degradation}.
In particular, for all attack types, the degradation curves peak near $12$~dBm and then decrease gradually as the transmit power increases. This behavior is directly reflected in the robustness index defined in \eqref{robustness}, which normalizes the rate loss under attack by the corresponding safe-system rate. Consequently, the peak degradation around $12$~dBm corresponds to the local minimum observed in the robustness index curves, while the subsequent reduction in degradation at higher transmit powers leads to the observed rebound of \ac{RSMA}. This rebound is consistent with the adaptive interference-management mechanism of \ac{RSMA}: as the impact of channel acquisition mismatch becomes more pronounced, the power allocation shifts toward the common message, mitigating the impact of inaccurate \ac{CSI} and reducing the relative rate gap to the safe case. In contrast, \ac{SDMA} lacks this adaptive mechanism, resulting in a more steadily decreasing robustness index under the same attack conditions.
Finally, Fig.~\ref{fig:res_3}(b) also confirms that the aligned attack with a fully connected \ac{RIS} yields the largest degradation among the considered architectures, which is consistent with the corresponding lower robustness levels observed in Fig.~\ref{fig:res_3}(a).

\subsection{Impact of RIS Architecture Parameters}\label{subsec:elements}

Figs.~\ref{fig:res_4}(a) and \ref{fig:res_4}(b) analyze the scaling behavior of the adversarial \ac{RIS} for a varying number of reflecting elements $D$ under the aligned interference attack, considering a fixed transmit power of $30$~dBm, while Fig.~\ref{fig:res_4}(c) examines the impact of the group size $D_g$ in the group-connected \ac{RIS} for a fixed number of reflecting elements.
As shown in Fig.~\ref{fig:res_4}(a), the average sum rate for both multiple access schemes degrades as the \ac{RIS} becomes larger. The performance of \ac{SDMA}, however, collapses at large $D$, particularly when attacked by a fully connected \ac{RIS}, whereas the decline of \ac{RSMA} is considerably less severe. For instance, at $D=400$, \ac{SDMA} falls below $12$~bits/s/Hz, while \ac{RSMA} still sustains above $16$~bits/s/Hz when both are attacked by the fully connected \ac{RIS}.

This trend is also observed in Fig.~\ref{fig:res_4}(b) in terms of sum-rate degradation, where the gap in degradation between \ac{RSMA} and \ac{SDMA} clearly widens as $D$ increases.
These results highlight the scaling advantage of the different \ac{RIS} architectures for an attacker, as the performance curves for the three architectures become more separated with larger $D$, demonstrating that \acp{BD-RIS} become increasingly advantageous over a conventional \ac{RIS}. At the same time, the results confirm the scalability of \ac{RSMA} in adversarial environments, since it continues to offer performance gains over \ac{SDMA} even when the adversarial \ac{RIS} dimension increases. Next, we examine how \ac{RSMA}'s robustness is affected by practical impairments.

Fig.~\ref{fig:res_4}(c) shows the effect of the group size $D_g$ in the group-connected \ac{RIS}, highlighting the tradeoff between implementation complexity and attack performance. For a fixed number of $D=200$ reflecting elements, increasing $D_g$ leads to a stronger attack and increased sum-rate degradation. Smaller group sizes result in lower performance and behavior closer to that of a single-connected \ac{RIS}, while benefiting from reduced hardware complexity and lower algorithmic complexity in the attack optimization, as discussed in Section~\ref{subsec:complexity}. In contrast, larger group sizes progressively approach the performance of a fully connected architecture at the cost of increased implementation and optimization complexity. These results confirm that the group-connected \ac{RIS} provides a flexible architecture that bridges the gap between single-connected and fully connected \ac{RIS} designs, allowing the adversary to balance complexity and attack effectiveness.

\subsection{Impact of CSI and SIC Imperfections}\label{subsec:imperfections}

\begin{figure}[t]
	\centering
	\includegraphics[width=1.0\linewidth]{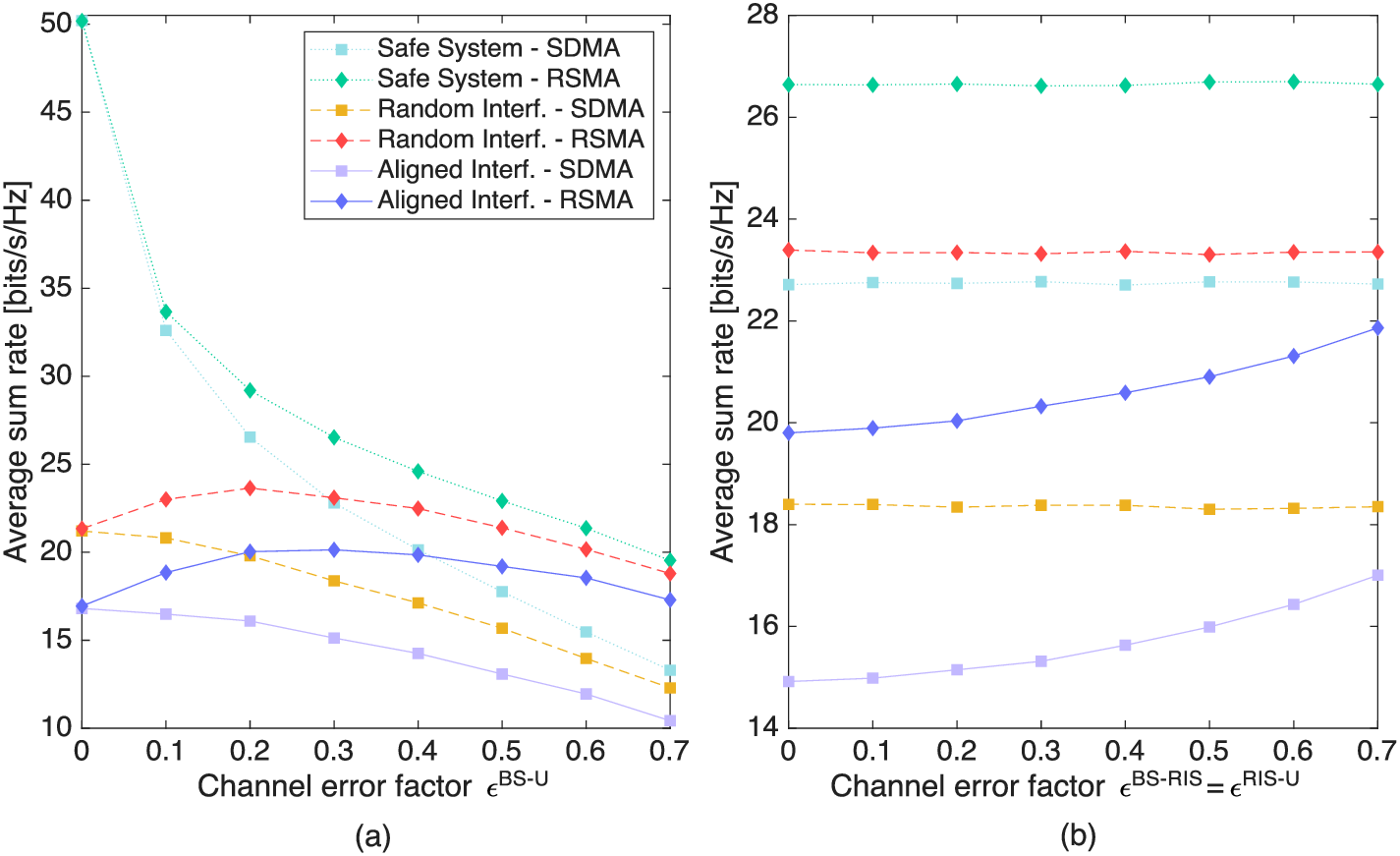}
	\caption{Impact of imperfect \ac{CSI} on the system sum rate: (a) versus channel error $\epsilon^{\text{\tiny BS-U}}$ in the BS-user link, with $\epsilon^{\text{\tiny BS-RIS}}=\epsilon^{\text{\tiny RIS-U}}=0.3$ for the \ac{RIS}-associated channels at the attacker; and (b) versus $\epsilon^{\text{\tiny BS-RIS}}=\epsilon^{\text{\tiny RIS-U}}$ at the attacker, with a fixed error of $\epsilon^{\text{\tiny BS-U}}=0.3$ in the BS-user link. Results are shown for both random and aligned attacks induced by the fully connected \ac{RIS}, with a transmit power of $30$~dBm.}\label{fig:res_5}
\end{figure}

Fig.~\ref{fig:res_5} provides a deeper analysis of the impact of imperfect \ac{CSI}, considering a fixed transmit power of $30$~dBm for both random and aligned attacks induced by the fully connected \ac{RIS}.
Fig.~\ref{fig:res_5}(a) investigates system performance versus the error in the legitimate BS-user link $\epsilon^{\text{\tiny BS-U}}$, while the attacker's \ac{CSI} for the \ac{RIS}-related links has a fixed error of $\epsilon^{\text{\tiny BS-RIS}} = \epsilon^{\text{\tiny RIS-U}} = 0.3$. As can be seen, at $\epsilon^{\text{\tiny RIS-U}}=0$, \ac{RSMA} and \ac{SDMA} exhibit identical performance under the same attack. This behavior is explained by the fact that \ac{RSMA} reduces to \ac{SDMA} when the \ac{BS} has perfect channel knowledge. As the error in the legitimate channel estimates increases, the sum rate of \ac{SDMA} under both attack types monotonically decreases. \ac{RSMA}, in contrast, shows a slight initial increase in its sum rate for small $\epsilon^{\text{\tiny BS-U}} > 0$ under both random and aligned attacks.

Fig.~\ref{fig:res_5}(b) examines the opposite scenario, where the legitimate BS-user \ac{CSI} error is fixed at $\epsilon^{\text{\tiny BS-U}}=0.3$, and the attacker's \ac{CSI} for the \ac{RIS}-related links is varied. The results show that as the attacker's channel estimation error increases, the performance of both \ac{RSMA} and \ac{SDMA} under attack improves. This is because higher estimation errors at the adversary lead to a less effective optimization for the aligned attack and a less coherent random channel for the random attack, resulting in less damaging interference. Across all scenarios where the legitimate channel is imperfect, i.e., $\epsilon^{\text{\tiny BS-U}} > 0$, \ac{RSMA} maintains a clear sum rate advantage over \ac{SDMA}.

\begin{figure}[t]
	\centering
	\includegraphics[width=1.0\linewidth]{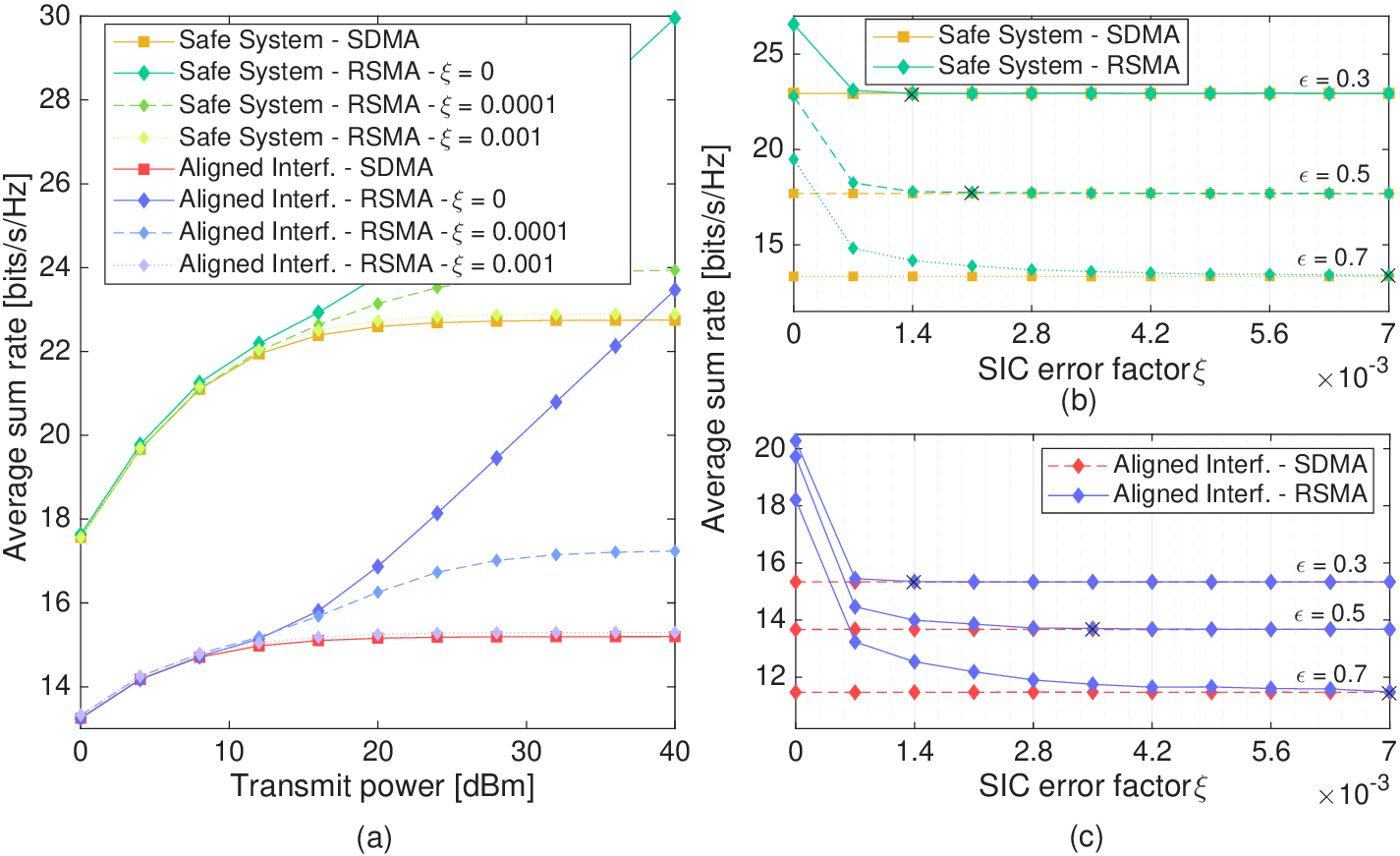}
    \caption{Impact of imperfect \ac{SIC} on the system sum rate: (a) versus transmit power for different \ac{SIC} error factors $\xi$, with all \ac{CSI} errors fixed to $0.3$; (b) versus the \ac{SIC} error factor $\xi$ under different \ac{CSI} error levels $\epsilon$, considering the safe scenario; and (c) versus the \ac{SIC} error factor $\xi$ under different \ac{CSI} error levels $\epsilon$ for the aligned interference attack. Subfigures (b) and (c) assume a fixed transmit power of $30$~dBm with $\epsilon \triangleq \epsilon^{\text{\tiny BS-U}} = \epsilon^{\text{\tiny BS-RIS}} = \epsilon^{\text{\tiny RIS-U}}$. The markers ``$\bm{\times}$'' indicate the smallest $\xi$ for which the RSMA and SDMA sum rate difference is lower than $0.1$~bits/s/Hz. Moreover, the aligned interference attack is induced by the fully connected \ac{RIS}.}
    \label{fig:res_6}
\end{figure}

Finally, Fig.~\ref{fig:res_6} investigates the system's sensitivity to \ac{SIC} imperfections under the aligned interference attack.
Fig.~\ref{fig:res_6}(a) illustrates the average sum rate versus transmit power for different \ac{SIC} error factors $\xi$. As expected, the performance of \ac{SDMA} is unaffected by this parameter. The results for \ac{RSMA}, however, reveal a high sensitivity to \ac{SIC} quality. Even a small error factor of $\xi = 10^{-4}$ causes a noticeable degradation compared to the perfect \ac{SIC} case ($\xi = 0$). For a larger error of $\xi = 10^{-3}$, the performance of \ac{RSMA} collapses, with its sum rate saturating at high power levels and nearly coinciding with that of \ac{SDMA}. This indicates that \ac{RSMA}'s ability to recover from the attack is highly dependent on the quality of the \ac{SIC} implementation.

The impact of imperfect \ac{SIC} is further detailed in Figs. \ref{fig:res_6}(b) and \ref{fig:res_6}(c) for a transmit power of $30$~dBm. The performance of \ac{RSMA} shows a steep decline as $\xi$ increases, in both the safe and attacked scenarios. In particular, the markers ``$\bm{\times}$'' in the figure indicate the data points at which the difference between the sum rates of RSMA and SDMA becomes lower than $0.1$~bits/s/Hz. Beyond these points, the robustness advantage of \ac{RSMA} vanishes, with the corresponding sum-rate curves practically overlapping. It is also noteworthy that this critical point depends on the \ac{CSI} error level $\epsilon = \epsilon^{\text{\tiny BS-U}} = \epsilon^{\text{\tiny BS-RIS}} = \epsilon^{\text{\tiny RIS-U}}$, as highlighted in the figures. For instance, both Figs.~\ref{fig:res_6}(b) and \ref{fig:res_6}(c) show that for larger $\epsilon$, the RSMA advantage is preserved over a wider range of $\xi$, despite an overall reduction in sum rate. This behavior reflects the different sensitivity of the common and private streams to \ac{CSI} errors, as well as the adopted power allocation strategy, which explicitly accounts for imperfect \ac{SIC} by incorporating the residual interference into the effective noise floor, allowing RSMA to at least match \ac{SDMA} under imperfect \ac{SIC}. These findings highlight a critical trade-off: while \ac{RSMA} is fundamentally more robust to external \ac{CSI} uncertainties, its overall performance advantage is contingent on the availability of high-quality hardware for interference cancellation.

Note that \ac{SDMA} was adopted as the baseline since it represents the dominant multi-user transmission strategy in current \ac{MIMO} systems and provides a natural benchmark for \ac{RSMA}, as they share the same spatial multiplexing strategy for private streams. Other multiple access schemes, such as \ac{OMA} and \ac{NOMA}, would also be affected by adversarial \ac{RIS}-induced imperfect \ac{CSI}. However, the detailed impact on these schemes depends on their specific transmission and receiver structures. For instance, in works that combine \ac{NOMA} with \ac{MIMO} systems, it is common to implement user clustering strategies and multiplex multiple user groups in the spatial domain. In such \ac{MIMO}--\ac{NOMA} systems, adversarial \ac{RIS}-induced channel manipulation could lead to increased inter-group interference. In \ac{OMA}-based \ac{MIMO} systems, where users are orthogonal in time or frequency, they can be served with dedicated precoders, and \ac{RIS}-induced attacks would impact system performance differently. In this case, the \ac{RIS}-corrupted \ac{CSI} would primarily degrade the alignment between the precoder and the true user channels, consequently reducing the \ac{BS} beamforming gain. A detailed investigation of these effects is beyond the scope of this work and is left for future research.

\section{Conclusions}
This paper has investigated the robustness of \ac{RSMA}-based multi-user \ac{MISO} systems to adversarial attacks induced by \acp{BD-RIS}. Two effective attack strategies were described and studied: random and aligned interference. For each, novel algorithms were proposed to generate valid reflection coefficients respecting the architectural \ac{BD-RIS} constraints. Extensive simulations revealed that \ac{RSMA} demonstrated significantly greater robustness to these attacks than the benchmark \ac{SDMA} scheme, especially under imperfect \ac{CSI}. Moreover, \ac{BD-RIS} significantly enhances adversarial efficacy compared to conventional \ac{RIS}, with particularly severe performance degradation observed when the \ac{BD-RIS} manipulates the uplink training phase via reflective operation.

The RSMA system considered in this work is modeled as being unaware of the presence of BD-RIS-adversarial manipulation.
Future work should expand this analysis and investigate the development of countermeasures against such threats to further enhance RSMA performance.
For example, a robust defense could involve the detection of RIS-induced attacks or anomalies based on uplink-downlink reciprocity checks and statistical monitoring of channel estimates. Once a threat is detected, the BS would proactively adapt the RSMA protocol by adjusting the rate-splitting structure and power allocation, as well as re-optimizing beamforming and channel estimation, thus mitigating the impact of adversarial attacks.
In parallel, complementary countermeasures could rely on making the uplink channel estimation phase less predictable through aperiodic or asynchronous training (similar to, e.g.,~\cite{darsena2022anti}), thus increasing the synchronization burden for the adversary.
In addition, further investigation of the adversarial potential of other emerging RIS concepts, such as multi-sector BD-RIS and non-reciprocal architectures, is also an interesting research direction.


\section*{Acknowledgments}
This work was supported by the Commonwealth Cyber Initiative (\href{https://www.cyberinitiative.org/}{cyberinitiative.org}) in Virginia, US, an investment to advance cyber R\&D, innovation, and workforce development, the National Science Foundation under grants no. 2326599 and 2318798, and the Research Council of Finland through the 6G-ConCoRSe project (grant no. 359850) and the 6G Flagship programme (grant no. 369116).

\ifCLASSOPTIONcaptionsoff
\newpage
\fi

\bibliographystyle{IEEEtran}
\bibliography{IEEEabbrv,main}

@article{Li2025NonReciprocalBDRIS,
  author  = {H. Li and B. Clerckx},
  title   = {{Non}-reciprocal beyond diagonal {RIS}: Multiport network models and performance benefits in full-duplex systems},
  journal = {IEEE Trans. Commun.},
  year    = {2025},
  volume  = {73},
  number  = {11},
  pages   = {12221--12234},
  month   = nov,
  doi     = {10.1109/TCOMM.2025.3568222}
}

@ARTICLE{Joudeh2016,
  author={H. {Joudeh} and B. {Clerckx}},
  journal={IEEE Trans. Signal Process.},
  title={Robust transmission in downlink multiuser {MISO} systems: A rate-splitting approach},
  year={2016},
  volume={64},
  number={23},
  pages={6227--6242},
  doi={10.1109/TSP.2016.2591501}
}

@inproceedings{Zhang2021SRSHRIS,
  author    = {H. Zhang and N. Shlezinger and I. Alamzadeh and G. C. Alexandropoulos and M. F. Imani and Y. C. Eldar},
  title     = {{Channel} estimation with simultaneous reflecting and sensing reconfigurable intelligent metasurfaces},
  booktitle = {Proc. IEEE 22nd Int. Workshop Signal Process. Adv. Wireless Commun. (SPAWC)},
  year      = {2021},
  pages     = {536--540},
  address   = {Lucca, Italy},
  month     = jun
}

@article{Katwe22,
  author  = {M. Katwe and K. Singh and B. Clerckx and C.-P. Li},
  title   = {{Rate}-splitting multiple access and dynamic user clustering for sum-rate maximization in multiple {RISs}-aided uplink {mmWave} system},
  journal = {IEEE Trans. Commun.},
  volume  = {70},
  number  = {11},
  pages   = {7365--7382},
  month   = nov,
  year    = {2022}
}

@article{Zhou23STAR,
  author  = {A. Dhok and S. Sharma},
  title   = {{Rate}-splitting multiple access with {STAR-RIS} over spatially-correlated channels},
  journal = {IEEE Trans. Commun.},
  volume  = {70},
  number  = {10},
  pages   = {6411--6427},
  month   = oct,
  year    = {2022}
}

@article{ISACRSMA24,
  author  = {H. Ke and J. Xu and W. Xu and C. Ding and D. W. K. Ng},
  title   = {{Rate} splitting and beamforming design for {RIS}-assisted {RSMA}-enhanced {ISAC} networks},
  journal = {IEEE Wireless Commun. Lett.},
  volume  = {14},
  number  = {6},
  pages   = {1738--1742},
  month   = jun,
  year    = {2025}
}

@article{shen2022scattering,
  author  = {S. Shen and B. Clerckx and R. Murch},
  title   = {{Modeling} and architecture design of reconfigurable intelligent surfaces using scattering parameter network analysis},
  journal = {IEEE Trans. Wireless Commun.},
  year    = {2022},
  volume  = {21},
  number  = {2},
  pages   = {1229--1243},
  doi     = {10.1109/TWC.2021.3103256}
}

@article{li2023bd_ris_modes,
  author  = {H. Li and S. Shen and B. Clerckx},
  title   = {{Beyond} diagonal reconfigurable intelligent surfaces: From transmitting and reflecting modes to single-, group-, and fully-connected architectures},
  journal = {IEEE Trans. Wireless Commun.},
  year    = {2023},
  volume  = {22},
  number  = {4},
  pages   = {2311--2324},
  doi     = {10.1109/TWC.2022.3210706}
}

@article{li2022nondiagonal_tvt,
  author  = {Q. Li and M. El-Hajjar and I. Hemadeh and A. Shojaeifard and A. A. M. Mourad and B. Clerckx and L. Hanzo},
  title   = {{Reconfigurable} intelligent surfaces relying on non-diagonal phase shift matrices},
  journal = {IEEE Trans. Veh. Technol.},
  year    = {2022},
  volume  = {71},
  number  = {6},
  pages   = {6367--6383},
  doi     = {10.1109/TVT.2022.3160364}
}

@article{fang2024lowcomplexity_bd,
  author  = {T. Fang and Y. Mao},
  title   = {{A} low-complexity beamforming design for beyond-diagonal {RIS}-aided multi-user networks},
  journal = {IEEE Commun. Lett.},
  year    = {2024},
  volume  = {28},
  number  = {1},
  pages   = {203--207},
  doi     = {10.1109/LCOMM.2023.3333411}
}

@inproceedings{fang2022fullyconnected_ris_rsma,
  author    = {T. Fang and Y. Mao and S. Shen and Z. Zhu and B. Clerckx},
  title     = {{Fully} connected reconfigurable intelligent surface aided rate-splitting multiple access for multi-user multi-antenna transmission},
  booktitle = {Proc. IEEE Int. Conf. Commun. Workshops (ICC Workshops)},
  year      = {2022},
  pages     = {675--680},
  doi       = {10.1109/ICCWorkshops53468.2022.9814551}
}

@article{li2024synergy_bd_ris_rsma,
  author  = {H. Li and S. Shen and B. Clerckx},
  title   = {{Synergizing} beyond diagonal reconfigurable intelligent surface and rate-splitting multiple access},
  journal = {IEEE Trans. Wireless Commun.},
  year    = {2024},
  volume  = {23},
  number  = {8},
  pages   = {8717--8729},
  doi     = {10.1109/TWC.2024.3353596}
}

@article{soleymani2024bd_ris_urllc,
  author  = {M. Soleymani and I. Santamaria and E. A. Jorswieck and B. Clerckx},
  title   = {{Optimization} of rate-splitting multiple access in beyond diagonal {RIS}-assisted {URLLC} systems},
  journal = {IEEE Trans. Wireless Commun.},
  year    = {2024},
  volume  = {23},
  number  = {5},
  pages   = {5063--5075},
  doi     = {10.1109/TWC.2023.3324190}
}

@inproceedings{khisa2025meta_uplink,
  author    = {S. Khisa and A. Amhaz and M. Elhattab and C. Assi and S. Sharafeddine},
  title     = {{Gradient}-based meta learning for uplink rate-splitting multiple access with beyond diagonal {RIS}},
  booktitle = {Proc. IEEE Int. Conf. Commun. (ICC)},
  year      = {2025},
  doi       = {10.1109/ICC52391.2025.11161088}
}

@article{asif2025rsma_swipt_bdris,
  author  = {M. Asif and Z. Ali and A. Ihsan and A. Ranjha and Z. Shoujin and M. Ahmed and X. Li and S. Chatzinotas},
  title   = {Robust Design of Beyond-Diagonal Reconfigurable Intelligent Surface Empowered {RSMA}-{SWIPT} System Under Channel Estimation Errors},
  journal = {arXiv},
  year    = {2025},
  note    = {arXiv:2508.08097}
}

@inproceedings{Iivanainen25,
  author    = {A. Iivanainen and R. Rajam{\"a}ki and V. Koivunen},
  title     = {{Beyond}-diagonal {RIS}: Adversarial channels and optimality of low-complexity architectures},
  booktitle = {Proc. IEEE 26th Int. Workshop Signal Process. Adv. Wireless Commun. (SPAWC)},
  year      = {2025}
}

@article{Halko2011,
  author  = {N. Halko and P.-G. Martinsson and J. A. Tropp},
  title   = {{Finding} structure with randomness: Probabilistic algorithms for constructing approximate matrix decompositions},
  journal = {SIAM Rev.},
  volume  = {53},
  number  = {2},
  pages   = {217--288},
  year    = {2011}
}

@article{Jin21,
  author  = {Y. Jin and J. Zhang and X. Zhang and H. Xiao and B. Ai and D. W. K. Ng},
  title   = {{Channel} estimation for semi-passive reconfigurable intelligent surfaces with enhanced deep residual networks},
  journal = {IEEE Trans. Veh. Technol.},
  year    = {2021},
  volume  = {70},
  number  = {10},
  pages   = {11083--11088}
}

@article{Akgun19,
  author  = {B. Akgun and M. Krunz and O. Ozan Koyluoglu},
  title   = {{Vulnerabilities} of massive {MIMO} systems to pilot contamination attacks},
  journal = {IEEE Trans. Inf. Forensics Security},
  year    = {2019},
  volume  = {14},
  number  = {5},
  pages   = {1251--1263}
}

@article{Sena20,
  author  = {A. S. de Sena and F. R. M. Lima and D. B. da Costa and Z. Ding and P. H. J. Nardelli and U. S. Dias and C. B. Papadias},
  title   = {{Massive} {MIMO}-{NOMA} networks with imperfect {SIC}: Design and fairness enhancement},
  journal = {IEEE Trans. Wireless Commun.},
  year    = {2020},
  volume  = {19},
  number  = {9},
  pages   = {6100--6115}
}

@article{Magnus80,
  author  = {J. R. Magnus and H. Neudecker},
  title   = {{The} elimination matrix: Some lemmas and applications},
  journal = {SIAM J. Algebraic Discrete Methods},
  volume  = {1},
  number  = {4},
  pages   = {422--449},
  year    = {1980}
}

@article{Dieci22,
  author  = {L. Dieci and A. Papini and A. Pugliese},
  title   = {{Takagi} factorization of matrices depending on parameters and locating degeneracies of singular values},
  journal = {SIAM J. Matrix Anal. Appl.},
  volume  = {43},
  number  = {3},
  pages   = {1148--1161},
  year    = {2022}
}

@inproceedings{staat2022mirror,
  author    = {P. Staat and H. Elders-Boll and M. Heinrichs and C. Zenger and C. Paar},
  title     = {{Mirror}, mirror on the wall: Wireless environment reconfiguration attacks based on fast software-controlled surfaces},
  booktitle = {Proc. Asia Conf. Comput. Commun. Secur.},
  year      = {2022},
  pages     = {208--221}
}

@article{wang2022wireless,
  author  = {Y. Wang and H. Lu and D. Zhao and Y. Deng and A. Nallanathan},
  title   = {{Wireless} communication in the presence of illegal reconfigurable intelligent surface: Signal leakage and interference attack},
  journal = {IEEE Wireless Commun.},
  year    = {2022},
  volume  = {29},
  number  = {3},
  pages   = {131--138}
}

@article{Wang24,
  author  = {H. Wang and Z. Han and A. L. Swindlehurst},
  title   = {{Channel} reciprocity attacks using intelligent surfaces with non-diagonal phase shifts},
  journal = {IEEE Open J. Commun. Soc.},
  year    = {2024},
  volume  = {5},
  pages   = {1469--1485},
  doi     = {10.1109/OJCOMS.2024.3370692}
}

@inproceedings{Sena24b,
  author    = {A. S. de Sena and A. Gomes and J. Kibilda and N. H. Mahmood and L. A. DaSilva and M. Latva-aho},
  title     = {{Malicious} {RIS} meets {RSMA}: Unveiling the robustness of rate splitting to {RIS}-induced attacks},
  booktitle = {Proc. IEEE Global Commun. Conf. (GLOBECOM)},
  year      = {2024},
  pages     = {3576--3581}
}

@article{Sena24,
  author  = {A. S. de Sena and J. Kibilda and N. H. Mahmood and A. Gomes and M. Latva-aho},
  title   = {{Malicious} {RIS} versus massive {MIMO}: Securing multiple access against {RIS}-based jamming attacks},
  journal = {IEEE Wireless Commun. Lett.},
  year    = {2024},
  volume  = {13},
  number  = {4},
  pages   = {989--993}
}

@article{Sena22,
  author  = {A. S. de Sena and P. H. J. Nardelli and D. B. da Costa and P. Popovski and C. B. Papadias},
  title   = {{Rate}-splitting multiple access and its interplay with intelligent reflecting surfaces},
  journal = {IEEE Commun. Mag.},
  year    = {2022},
  volume  = {60},
  number  = {7},
  pages   = {52--57}
}

@article{Konar17,
  author  = {A. Konar and N. D. Sidiropoulos},
  title   = {{Fast} approximation algorithms for a class of non-convex {QCQP} problems using first-order methods},
  journal = {IEEE Trans. Signal Process.},
  year    = {2017},
  volume  = {65},
  number  = {13},
  pages   = {3494--3509}
}

@article{lyu2020irs,
  author  = {B. Lyu and D. T. Hoang and S. Gong and D. Niyato and D. I. Kim},
  title   = {{IRS}-based wireless jamming attacks: When jammers can attack without power},
  journal = {IEEE Wireless Commun. Lett.},
  year    = {2020},
  volume  = {9},
  number  = {10},
  pages   = {1663--1667}
}

@article{huang2023disco,
  author  = {H. Huang and Y. Zhang and H. Zhang and Y. Cai and A. L. Swindlehurst and Z. Han},
  title   = {{Disco} intelligent reflecting surfaces: Active channel aging for fully-passive jamming attacks},
  journal = {IEEE Trans. Wireless Commun.},
  year    = {2024},
  volume  = {23},
  number  = {1},
  pages   = {806--819}
}

@article{li2023reconfigurable,
  author  = {H. Li and S. Shen and M. Nerini and B. Clerckx},
  title   = {{Reconfigurable} intelligent surfaces 2.0: Beyond diagonal phase shift matrices},
  journal = {IEEE Commun. Mag.},
  year    = {2023},
  volume  = {62},
  number  = {3},
  pages   = {102--108}
}

@article{reifert2023comeback,
  author  = {R.{-}J. Reifert and S. Roth and A. A. Ahmad and A. Sezgin},
  title   = {{Comeback} kid: Resilience for mixed-critical wireless network resource management},
  journal = {IEEE Trans. Veh. Technol.},
  year    = {2023},
  volume  = {72},
  number  = {12},
  pages   = {16177--16194}
}

@article{rivetti2024malicious,
  author  = {S. Rivetti and O. T. Demir and E. Bj{\"o}rnson and M. Skoglund},
  title   = {{Malicious} reconfigurable intelligent surfaces: How impactful can destructive beamforming be?},
  journal = {IEEE Wireless Commun. Lett.},
  year    = {2024},
  volume  = {13},
  number  = {7},
  pages   = {1918--1922},
  doi     = {10.1109/LWC.2024.3395831}
}

@article{alakoca2023metasurface,
  author  = {H. Alakoca and M. Namdar and S. Aldirmaz-Colak and M. Basaran and A. Basgumus and L. Durak-Ata and H. Yanikomeroglu},
  title   = {{Metasurface} manipulation attacks: Potential security threats of {RIS}-aided {6G} communications},
  journal = {IEEE Commun. Mag.},
  year    = {2023},
  volume  = {61},
  number  = {1},
  pages   = {24--30},
  doi     = {10.1109/MCOM.005.2200162}
}

@article{huang2024disco,
  author  = {H. Huang and L. Dai and H. Zhang and C. Zhang and Z. Tian and Y. Cai and A. L. Swindlehurst and Z. Han},
  title   = {{DISCO} might not be funky: Random intelligent reflective surface configurations that attack},
  journal = {IEEE Wireless Commun.},
  year    = {2024},
  volume  = {31},
  number  = {5},
  pages   = {76--82},
  doi     = {10.1109/MWC.014.2300470}
}

@article{huang2024disco-omni,
  author  = {H. Huang and H. Zhang and J. Yuan and L. Sun and Y. Wang and W. Mei and B. Di and Y. Cai and Z. Han},
  title   = {{Disco} intelligent omni-surfaces: 360-degree fully-passive jamming attacks},
  journal = {IEEE Trans. Wireless Commun.},
  year    = {2026},
  volume  = {25},
  pages   = {61--74},
  doi     = {10.1109/TWC.2025.3581208}
}

@inproceedings{wang2024beyond,
  author    = {H. Wang and J. Nossek and A. L. Swindlehurst},
  title     = {{Beyond}-diagonal {RIS} attacks on physical layer key generation},
  booktitle = {Proc. IEEE Int. Workshop Signal Process. Adv. Wireless Commun. (SPAWC)},
  year      = {2024},
  pages     = {946--950},
  doi       = {10.1109/SPAWC60668.2024.10694006}
}

@article{huang2024integrated,
  author  = {H. Huang and H. Zhang and W. Mei and J. Li and Y. Cai and A. L. Swindlehurst and Z. Han},
  title   = {{Integrated} sensing and communication under {DISCO} physical-layer jamming attacks},
  journal = {IEEE Wireless Commun. Lett.},
  year    = {2024},
  volume  = {13},
  number  = {11},
  pages   = {3044--3048},
  doi     = {10.1109/LWC.2024.3439398}
}

@inproceedings{wang2024non-diagonal,
  author    = {H. Wang and Z. Han and A. L. Swindlehurst},
  title     = {{Non}-diagonal {RIS} empowered channel reciprocity attacks on {TDD}-based wireless systems},
  booktitle = {Proc. IEEE Int. Conf. Commun. (ICC)},
  year      = {2024},
  pages     = {127--132},
  doi       = {10.1109/ICC51166.2024.10622674}
}

@inproceedings{gomes2024beam,
  author    = {A. Gomes and A. S. de Sena and N. H. Mahmood and M. Latva-aho and L. A. DaSilva and J. Kibi{\l}da},
  title     = {{Beam} management manipulation with adversarial reconfigurable intelligent surfaces},
  booktitle = {Proc. IEEE Global Commun. Conf. (GLOBECOM)},
  year      = {2024},
  pages     = {3231--3236},
  doi       = {10.1109/GLOBECOM52923.2024.10900964}
}

@article{li2024ris-jamming,
  author  = {G. Li and P. Staat and H. Li and M. Heinrichs and C. Zenger and R. Kronberger and H. Elders-Boll and C. Paar and A. Hu},
  title   = {{RIS}-jamming: Breaking key consistency in channel reciprocity-based key generation},
  journal = {IEEE Trans. Inf. Forensics Security},
  year    = {2024},
  volume  = {19},
  pages   = {5090--5105},
  doi     = {10.1109/TIFS.2024.3389569}
}

@inproceedings{rivetti2025destructive,
  author    = {S. Rivetti and {\"O}. T. Demir and E. Bj{\"o}rnson and M. Skoglund},
  title     = {{Destructive} and constructive {RIS} beamforming in an {ISAC} multi-user {MIMO} network},
  booktitle = {Proc. IEEE Int. Conf. Commun. (ICC)},
  year      = {2025},
  pages     = {2412--2417},
  doi       = {10.1109/ICC52391.2025.11162058}
}

@article{clerckx2023primer,
  author  = {B. Clerckx and Y. Mao and E. A. Jorswieck and J. Yuan and D. J. Love and E. Erkip and D. Niyato},
  title   = {{A} primer on rate-splitting multiple access: Tutorial, myths, and frequently asked questions},
  journal = {IEEE J. Sel. Areas Commun.},
  year    = {2023},
  volume  = {41},
  number  = {5},
  pages   = {1265--1308},
  doi     = {10.1109/JSAC.2023.3242718}
}

@article{xu2022rate,
  author  = {Y. Xu and Y. Mao and O. Dizdar and B. Clerckx},
  title   = {{Rate}-splitting multiple access with finite blocklength for short-packet and low-latency downlink communications},
  journal = {IEEE Trans. Veh. Technol.},
  year    = {2022},
  volume  = {71},
  number  = {11},
  pages   = {12333--12337},
  doi     = {10.1109/TVT.2022.3191085}
}

@article{lu2024outage,
  author  = {H. Lu and X. Xie and Z. Shi and H. Lei and N. Zhao and J. Cai},
  title   = {{Outage} performance of uplink rate splitting multiple access with randomly deployed users},
  journal = {IEEE Trans. Wireless Commun.},
  year    = {2024},
  volume  = {23},
  number  = {2},
  pages   = {1308--1326},
  doi     = {10.1109/TWC.2023.3288110}
}

@article{dizdar2021rate,
  author  = {O. Dizdar and Y. Mao and B. Clerckx},
  title   = {{Rate}-splitting multiple access to mitigate the curse of mobility in (massive) {MIMO} networks},
  journal = {IEEE Trans. Commun.},
  year    = {2021},
  volume  = {69},
  number  = {10},
  pages   = {6765--6780},
  doi     = {10.1109/TCOMM.2021.3098695}
}

@article{pala2-24spectral,
  author  = {S. Pala and M. Katwe and K. Singh and B. Clerckx and C.-P. Li},
  title   = {{Spectral}-efficient {RIS}-aided {RSMA} {URLLC}: Toward mobile broadband reliable low latency communication ({mBRLLC}) system},
  journal = {IEEE Trans. Wireless Commun.},
  year    = {2024},
  volume  = {23},
  number  = {4},
  pages   = {3507--3524},
  doi     = {10.1109/TWC.2023.3309028}
}

@article{zhou2021rate,
  author  = {G. Zhou and Y. Mao and B. Clerckx},
  title   = {{Rate}-splitting multiple access for multi-antenna downlink communication systems: Spectral and energy efficiency tradeoff},
  journal = {IEEE Trans. Wireless Commun.},
  year    = {2022},
  volume  = {21},
  number  = {7},
  pages   = {4816--4828},
  doi     = {10.1109/TWC.2021.3133433}
}

@article{darsena2022anti,
  title={Anti-jamming beam alignment in millimeter-wave {MIMO} systems},
  author={Darsena, Donatella and Verde, Francesco},
  journal={IEEE Trans. Commun.},
  volume={70},
  number={8},
  pages={5417--5433},
  year={2022},
  publisher={IEEE}
}

@article{zheng2024mutual,
  title={{Mutual coupling in {RIS}-aided communication: {Model} training and experimental validation}},
  author={Zheng, Pinjun and Wang, Ruiqi and Shamim, Atif and Al-Naffouri, Tareq Y},
  journal={IEEE Trans. Wireless Commun.},
  year={2024},
  publisher={IEEE}
}

@inproceedings{de2025performance,
  author    = {J. C. da Silva Filho and J. V. de Ara{\'u}jo and B. Sokal and A. L. F. de Almeida},
  title     = {Performance Evaluation of Beyond Diagonal {RIS} under Hardware Impairments},
  booktitle = {Proc. Brazilian Symp. Telecommun. Signal Process. (SBrT)},
  address   = {Natal, Brazil},
  month     = sep,
  year      = {2025}
}

@article{Li2024RHS_HWI,
  author  = {Q. Li and M. El-Hajjar and Y. Sun and L. Hanzo},
  title   = {Performance Analysis of Reconfigurable Holographic Surfaces in the Near-Field Scenario of Cell-Free Networks Under Hardware Impairments},
  journal = {IEEE Trans. Wireless Commun.},
  volume  = {23},
  number  = {9},
  pages   = {11972--11984},
  month   = sep,
  year    = {2024}
}

\vspace{-3mm}

\begin{IEEEbiography}[{\includegraphics[width=1in,height=1.25in,clip,keepaspectratio]{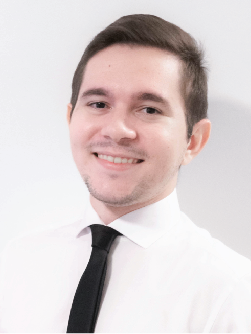}}]{Arthur Sousa de Sena} 

(Member, IEEE) received his B.Sc. degree in Computer Engineering from the Federal University of Ceará (UFC), Brazil, in 2017, with an exchange period at Illinois Institute of Technology, USA, from August 2014 to December 2015. He received his M.Sc. degree in Teleinformatics Engineering, also from UFC, in 2019, and his D.Sc. degree (with distinction) in Electrical Engineering from LUT University, Finland, in 2022.
Currently, Dr. Sena is a Postdoctoral Researcher at the Centre for Wireless Communication at the University of Oulu, Finland, where he leads multiple research projects. Previously, he was a Researcher at the AI and Digital Science Research Center at the Technology Innovation Institute, Abu Dhabi, UAE, from November 2022 to May 2023. From 2019 to 2022, he was a Junior Researcher in the Cyber-Physical Systems Group at LUT University.
His research interests span the broad areas of wireless communications and signal processing.
He received the Nokia Foundation Award in October 2020, the LUT Research Foundation Award in December 2020, and the IEEE Global Communications Conference (GLOBECOM) Best Paper Award in December 2022. He has authored several peer-reviewed papers in prestigious journals and flagship conferences. He serves as an Editor for IEEE Communications Letters and is a member of the IEEE Communications Society.
\end{IEEEbiography}

\begin{IEEEbiography}[{\includegraphics[width=1in,height=1.25in,clip,keepaspectratio]{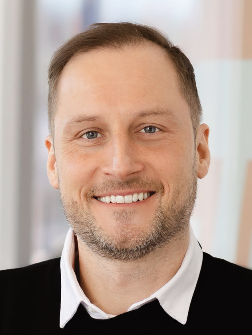}}]{Jacek Kibi\l{}da}

(Senior Member, IEEE) is an Associate Professor with the Commonwealth Cyber Initiative and the Department of Electrical and Computer Engineering at Virginia Tech. Jacek received his Ph.D. degree from Trinity College Dublin and his M.Sc. from Poznan University of Technology. Jacek's research focuses on reliability, resilience, and robustness in wireless communications and networks. He is a recipient of the Best Paper Award at IEEE GLOBECOM 2023 and the Best Demo Award at IEEE MILCOM 2023.

\end{IEEEbiography}

\begin{IEEEbiography}[{\includegraphics[width=1in,height=1.25in,clip,keepaspectratio]{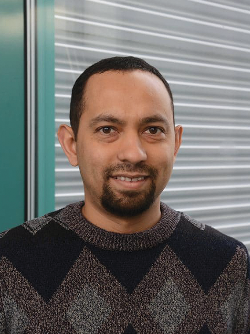}}]{Nurul Huda Mahmood}

(Member, IEEE) was born in Bangladesh. He received the Ph.D. degree from NTNU, Norway, in 2013. He is currently a Senior Research Fellow with CWC, University of Oulu, Finland, and the coordinator of wireless connectivity research in the Finnish 6G Flagship program. Before joining the University of Oulu in 2018, he was an Associate Professor with the Department of Electronics Systems at Aalborg University, Denmark. Nurul has contributed to various international research projects, most recently as a WP leader in Hexa-X-II-EU’s flagship 6G research project. He has coauthored over 100 peer-reviewed publications. His current research interests include resilient communications for wireless networks.

\end{IEEEbiography}

\begin{IEEEbiography}[{\includegraphics[width=1in,height=1.25in,clip,keepaspectratio]{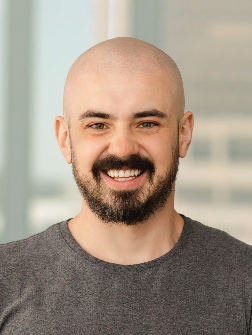}}]{André Gomes}

(Member, IEEE) is a tenure-track Assistant Professor of Computer Science at Rowan University, US. He received his PhD in Computer Engineering from Virginia Tech, MS in Computer Science from Universidade Federal de Minas Gerais, and BS in Telecommunications Engineering from Universidade Federal de São João del-Rei. He was also a Postdoctoral Associate with CCI, US, before joining Rowan. His areas of interest and research relate to wireless networking, with a focus on reliability, resilience, and controllability. He is a recipient of a best paper award at IEEE GLOBECOM 2023.

\end{IEEEbiography}

\begin{IEEEbiography}[{\includegraphics[width=1in,height=1.25in,clip,keepaspectratio]{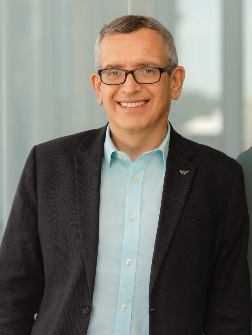}}]{Luiz DaSilva}

(Fellow, IEEE) is the Bradley Professor of Cybersecurity at Virginia Tech, where he serves as the inaugural Executive Director of the Commonwealth Cyber Initiative, a consortium of 47 institutions of higher education in Virginia, USA, with a mission of research, innovation, and workforce development in cybersecurity. Prof. DaSilva previously held the Chair of Telecommunications at Trinity College Dublin, Ireland. He is the author of more than 300 peer-reviewed publications and a Fellow of the IEEE for contributions to cognitive networks and wireless resource management. He has also been a Fellow of Trinity College Dublin, a Distinguished Lecturer of the IEEE Communication Society, and a Virginia Tech College of Engineering Faculty Fellow.

\end{IEEEbiography}

\begin{IEEEbiography}[{\includegraphics[width=1in,height=1.25in,clip,keepaspectratio]{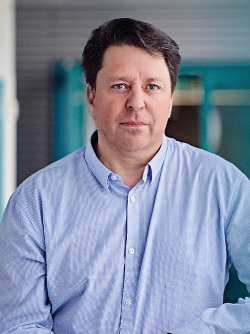}}]{Matti Latva-aho}

(Fellow, IEEE) is a distinguished expert in wireless communications. He holds M.Sc., Lic.Tech., and Dr.Tech. (Hons.) degrees in Electrical Engineering from the University of Oulu, Finland, awarded in 1992, 1996, and 1998, respectively. From 1992 to 1993, he worked as a Research Engineer at Nokia Mobile Phones in Oulu before joining the Centre for Wireless Communications (CWC) at the University of Oulu. Prof. Latva-aho served as Director of CWC from 1998 to 2006 and later as Head of the Department of Communication Engineering until August 2014. He was nominated as an Academy Professor by the Academy of Finland in 2017. He is a Professor of Wireless Communications at the University of Oulu and served as Director of the National 6G Flagship Programme. He is also a Global Fellow at The University of Tokyo. In 2025, he was appointed Vice-Rector for Research at the University of Oulu for a five-year term. With an extensive portfolio of over 600 conference and journal publications, Prof. Latva-aho has significantly advanced the field of wireless communications. His contributions were recognized in 2015 when he received the prestigious Nokia Foundation Award for his groundbreaking research in mobile communications.

\end{IEEEbiography}

\end{document}